\documentclass[twocolumn,superscriptaddress]{revtex4-1}
\usepackage{amsmath,bm}
\usepackage{color}
\usepackage{graphicx}

\usepackage[whole]{bxcjkjatype} 
\usepackage[colorlinks=true,urlcolor=blue,citecolor=blue,linkcolor=blue,breaklinks=true]{hyperref}
\input glyphtounicode
\usepackage{txfonts}

\begin{document}
\title{Current response of nonequilibrium steady states in Landau-Zener problem:\\
Nonequilibrium Green's function approach}
\author{Sota Kitamura}
\affiliation{Department of Applied Physics, The University of Tokyo, Hongo, Tokyo,
113-8656, Japan}
\author{Naoto Nagaosa}
\affiliation{Department of Applied Physics, The University of Tokyo, Hongo, Tokyo,
113-8656, Japan}
\affiliation{RIKEN Center for Emergent Matter Science (CEMS), Wako, Saitama, 351-0198,
Japan}
\author{Takahiro Morimoto}
\affiliation{Department of Applied Physics, The University of Tokyo, Hongo, Tokyo,
113-8656, Japan}
\affiliation{JST, PRESTO, Kawaguchi, Saitama, 332-0012, Japan}
\date{\today}
\begin{abstract}
The carrier generation in insulators subjected to strong electric fields is characterized by the Landau-Zener formula for the tunneling probability with a nonperturbative exponent. 
Despite its long history with diverse applications and extensions, study of nonequilibrium steady states and associated current response in the presence of the generated carriers has been mainly limited to numerical simulations so far.
Here, we develop a framework to calculate the nonequilibrium Green's function of generic insulating systems under a DC electric field, in the presence of a fermionic reservoir. Using asymptotic expansion techniques, we derive a semi-quantitative formula for the Green's function with nonperturbative contribution. This formalism enables us to calculate dissipative current response of the nonequilibrium steady state, which turns out to be not simply characterized by the intraband current proportional to the tunneling probability. We also apply the present formalism to noncentrosymmetric insulators, and propose nonreciprocal charge and spin transport peculiar to  tunneling electrons.
\end{abstract}
\maketitle

\section{Introduction}

Nonperturbative effects, which cannot be captured by order-by-order
calculation, lead to a drastic change in the property of materials.
The Landau-Zener tunneling~\cite{LandauLifshitz,Zener32} is a representative nonperturbative phenomenon,
where application of an intense electric field to insulators leads
to a rapid increase in the carrier generation rate. 

Responses of quantum materials against external stimuli show a rich
variety according to the symmetries of the underlying microscopic Hamiltonian.
In particular, nonreciprocal transport is an important class of phenomena
extensively explored both in linear and nonlinear regime~\cite{Tokura-Nagaosa18,Rikken01,RikkenCNT,RikkenNature,RikkenSi,Pop14,Wakatsuki17,Yasuda20}. While the
nonreciprocal response with a directional transport requires broken
inversion symmetry, the presence of the time-reversal symmetry sometimes
forbids the directionality, as typified in Onsager's reciprocal relation
on generic linear responses~\cite{Onsager}. 

Recent developments on the study of the nonlinear responses with a
topological/geometric origin~\cite{Sipe,Young-Rappe,Cook17,Morimoto-Nagaosa16,Nagaosa-Morimoto17,Nagaosa20} suggest that the nonperturbative regime
also hosts diverse novel phenomena including nonreciprocal transport
and topological responses. Indeed, the nonreciprocity in the tunneling
probability due to the geometric phase effect has been proposed recently~\cite{Kitamura2020,Takayoshi20}. 

Despite the potential importance, transport properties in the nonperturbative
regime have not been explored so intensively.  
For the tunneling problems, quantitative estimation of the electric
current associated with the tunneling carriers in the nonequilibrium states has been missing, except for several numerical studies in graphene~\cite{Barreiro2009,Vandecasteele2010,Fang2011,Li2018} and correlated insulators~\cite{Oka2003,Okamoto2007,Sugimoto2008,Eckstein2010,Heidrich2010,Tsuji08},
although the tunneling probability in the equilibrium (or in a mesoscopic environment) has been studied in a broad context~\cite{Davis1976,Dykhne1962,George1974,Kayanuma1984,Berry1990,Joye1991-1,Joye1991-2,Ao89,Ao1991,Liu02,Saito07,Kayanuma08,Oka2012}.
The difficulty to do so stems from the far from equilibrium  nature of the distribution of the excited electrons in the nonperturbative regime. To determine the
nonequilibrium steady state, we have to deal with the Green's function
or density matrix of the system in an open-dissipative setup. 
While such methods with the nonequilibrium ensemble are actively
studied~\cite{Rammer86,Buttiker86,Jauho94,Aoki2014,Gorini1976,Lindblad1976,Gisin1992,Breuer2002}, it is still a nontrivial problem how to incorporate such nonequilibrium nature
with the nonperturbative treatment of the tunneling process in the
wave-function based theory.

\begin{figure}
\centering{}\includegraphics[width=.95\linewidth]{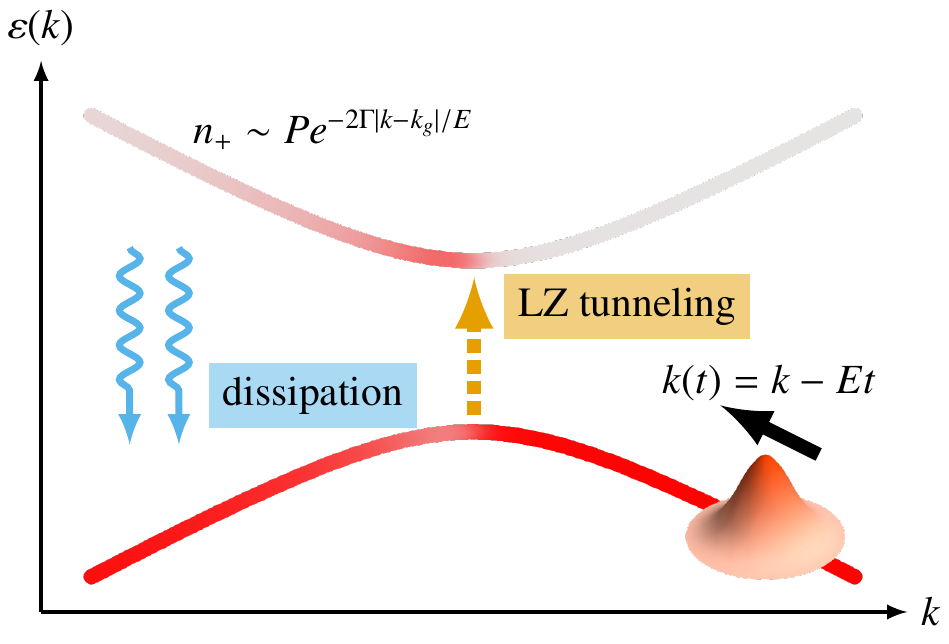}\caption{\label{fig:schematic}Schematic picture of the nonequilibrium steady state for the open-dissipative Landau-Zener problem.
The energy dispersion of a two-band insulator is colored in red, according to the occupation number $n_\pm$.
Electrons driven by a static electric field $E$ undergo the Landau-Zener tunneling with a probability $P$, when passing through the gap minimum. Excited electrons have a lifetime $t\sim1/2\Gamma$ due to the coupling to the fermionic reservoir, which results in the exponential decay of the momentum distribution.}
\end{figure}

In this paper, we consider a band insulator coupled to a fermionic
particle reservoir under a DC electric field. 
The nonequilbirum steady state of this setup, schematically depicted in Fig.~\ref{fig:schematic}, is realized as a result of a balance between the nonperturbative excitation and relaxation due to the dissipation. 
We derive a concise formula for the
nonequilibrium Green's function of the steady state, 
which includes a contribution from the nonperturbative tunneling process as well as the dissipative effect. 
This enables us to study
the electric current due to the excited electrons, which exhibits nontrivial
behaviors which cannot be deduced from the property
of the tunneling probability. We clarify that there appears a competition
between intraband and interband current, which have different dependence
on the electric field. We also apply the obtained formula to noncentrosymmetric
insulators, in order to discuss the nonreciprocal transport. We reveal
novel phenomena, i.e., a crossover of the nonreciprocity ratio due to
the competition mentioned above, and the nonreciprocal spin current
due to the asymmetric band dispersion. Such nonreciprocal spin current of tunneling electrons may be related to chiral-induced spin selectivity (CISS) found in DNA molecules, where photoexcited electrons show spin accumulation through propagating in insulating DNA molecules~\cite{Gohler11,Matityahu16}.

This paper is organized as follows. In Sec.~\ref{sec:Formulation},
we develop a framework to calculate the nonequilibrium Green's function
of the tunneling problem. We first review the calculation of the tunneling
probability in isolated systems in Sec.~\ref{subsec:Tunneling-probability}.
We introduce a key method, the adiabatic perturbation theory here. 
We extend this framework to open systems
in Sec.~\ref{subsec:Nonequilibrium-Green's-function}, and construct
the nonequilibrium Green's function using the solution of the equation
of motion for the isolated system. We show the numerically-calculated carrier density of the open system using the proposed framework in Sec.~\ref{subsec:Numerical-calculation}.
We perform an asymptotic expansion
for the nonequilibrium Green's function in Sec.~\ref{sec:Asymptotic-evaluation},
in order to derive approximate analytic expressions. 
We summarize the main results in Sec.~\ref{subsec:Overview} with a brief sketch
of the derivation. 
We provide detail of the derivation with starting from the adiabatic limit in Sec.~\ref{subsec:quasiequilibrium}, 
where we find that the asymptotic evaluation reproduces the result of the
Boltzmann equation with the relaxation-time approximation. We combine
this with the method of the contour integral, to obtain the nonperturbative
correction to the Green's function, in Sec.~\ref{subsec:Tunneling-contribution}.
We discuss the application of the obtained formula in Sec.~\ref{sec:Applications}.
We discuss the nonperturbative electric current and associated nonreciprocity,
as well as the extension of the formalism to lattice systems. Finally,
we conclude the paper in Sec.~\ref{sec:Conclusion}.

\section{Formulation\label{sec:Formulation}}

\subsection{Tunneling probability\label{subsec:Tunneling-probability}}

We start with reviewing how the tunneling probability is described
in isolated systems. The open-system formalism will be developed in
the next subsection, based on the approach taken here. 

In calculating the tunneling probability, the adiabatic perturbation
theory~\cite{Dykhne1962,Davis1976,George1974,DeGrandi2010,Kitamura2020}, a series expansion with respect to a slowly changing parameter,
plays a key role in capturing the nonperturbative nature. To see this,
let us introduce a $2\times2$ Hamiltonian $H$ in the momentum space
(in the first-quantized form), 
\begin{equation}
H(k)|u_{\pm,k}\rangle=\varepsilon_{\pm}(k)|u_{\pm,k}\rangle,\label{eq:hamiltonian-isolated}
\end{equation}
and consider its adiabatic time evolution. Here, $|u_{\alpha,k}\rangle$
is the Bloch wave function of the upper ($\alpha=+$) and lower ($\alpha=-$)
band with crystal momentum $k$ and eigenenergy $\varepsilon_{\alpha}(k)$.
In this study we consider a gapped case, $\varepsilon_{-}(k)<\varepsilon_{+}(k)$. 

We introduce a DC electric field $E$ via the Peierls substitution,
$H(k)\rightarrow H(k-Et)$, where we set $e=\hbar=1$ for simplicity. We consider the time
evolution described by the time-dependent Schr\"{o}dinger equation,
\begin{equation}
i\partial_{t}|\Phi(t)\rangle=H(k-Et)|\Phi(t)\rangle.
\end{equation}
We set the initial state at $t=t_i\rightarrow-\infty$ to be the
eigenstate on the lower band, i.e., $|\Phi(t_i)\rangle=|\psi_{-,k}(t_i)\rangle\propto|u_{-,k-Et_i}\rangle$ [See Eq.~(\ref{eq:snapshot-basis}) below].

It is well-known as the adiabatic theorem that $|\langle u_{-,k-Et}|\Phi(t)\rangle|^{2}\rightarrow1$
in the weak field limit $E\rightarrow0$. The tunneling probability,
i.e. the probability to observe the state in the upper band (usually
after a long time), 
\begin{equation}
P=|\langle u_{+,k-Et}|\Phi(t)\rangle|^{2}=1-|\langle u_{-,k-Et}|\Phi(t)\rangle|^{2},
\end{equation}
 thus measures how much the adiabatic theorem is violated due to nonzero
field strength $E\neq0$. While this observation implies that it is
convenient to expand $|\Phi(t)\rangle$ into the snapshot eigenstates
$|u_{\pm,k-Et}\rangle$, we here introduce a suitable basis with an
additional phase factor,
\begin{equation}
|\psi_{\alpha,k}(t)\rangle=e^{-i\int_{t_0}^{t}dt^{\prime}(\varepsilon_{\alpha}(k-Et^{\prime})+EA_{\alpha\alpha}(k-Et^{\prime}))}|u_{\alpha,k-Et}\rangle,\label{eq:snapshot-basis}
\end{equation}
where $A_{\alpha\beta}(k)=i\langle u_{\alpha,k}|\partial_{k}|u_{\beta,k}\rangle$
is the Berry connection. Note that the lower limit of the $t^{\prime}$
integral is chosen to $t_0\coloneqq k/E\neq t_i$ for future convenience. 
Hereafter we omit the arguments $k-Et$ when
it is not confusing. While $|u_{\alpha,k-Et}\rangle$ is not necessarily
smooth because of the arbitrariness of the phase factor (as a function
of $k$), $|\psi_{\alpha,k}(t)\rangle$ does not depend on a gauge choice of $|u_{\alpha,k-Et}\rangle$~\footnote{{The gauge transformation we consider here is defined as 
$|u_{\alpha,k}\rangle\rightarrow|u_{\alpha,k}\rangle e^{i\Lambda_\alpha(k)}$ with an arbitrary real function $\Lambda_\alpha$. 
Since the Berry connection is transformed as $A_{\alpha\alpha}(k)\rightarrow A_{\alpha\alpha}(k)-\partial_k \Lambda_\alpha(k)$, 
one can check that indeed the snapshot basis does not depend on the gauge choice of $|u_{\alpha,k}\rangle$ (except for the overall time-independent phase factor arising from the gauge choice at the initial time, i.e., $|\psi_{\alpha,k}(t)\rangle\rightarrow |\psi_{\alpha,k}(t)\rangle e^{i\Lambda_\alpha(k-Et_0)}$).}} and is a smooth
function of $t$, thanks to the Berry phase factor. We call $|\psi_{\alpha,k}(t)\rangle$
the snapshot basis throughout this paper. 

Now, by expanding $|\Phi(t)\rangle$ as
\begin{equation}
|\Phi(t)\rangle=\sum_{\alpha=\pm}a_{\alpha}(t)|\psi_{\alpha,k}(t)\rangle
\end{equation}
with $a_{-}(-\infty)=1$ and $a_{+}(-\infty)=0$, we obtain the equation
of motion for $a_{\pm}(t)$ as 

\begin{equation}
i\begin{pmatrix}\dot{a}_{+}(t)\\
\dot{a}_{-}(t)
\end{pmatrix}=\begin{pmatrix}0 & W(t)\\
W^{\ast}(t) & 0
\end{pmatrix}\begin{pmatrix}a_{+}(t)\\
a_{-}(t)
\end{pmatrix},\label{eq:eom-ap}
\end{equation}
where 
\begin{equation}
W(t)=EA_{+-}(k-Et)e^{i\int_{t_0}^{t}dt^{\prime}(\varepsilon_{+}-\varepsilon_{-}+E(A_{++}-A_{--}))}.
\label{eq:Wt}
\end{equation}
The adiabatic theorem immediately follows from the fact that $W(t)\rightarrow0$
as $E\rightarrow0$. 

As $|W(t)|=o(|E|^{0})$, we can regard $W(t)$ as a perturbation to
the adiabatic time evolution. Within the first-order, we obtain 
\begin{align}
a_{+}(t) & \simeq-i\int_{-\infty}^{t}dt_{1}W(t_{1}).\label{eq:ap1st}
\end{align}
The formal full solution can also be obtained using the time-ordered
exponential. The tunneling probability is now evaluated as $P=|a_{+}(t)|^{2}$.

As is well-known as the Dykhne-Davis-Pechukas (DDP) method~\cite{Dykhne1962,Davis1976}, in $t\rightarrow\infty$,
one can evaluate Eq.~(\ref{eq:ap1st}) asymptotically by employing
the contour integral in the complexified $t_1$ plane, which yields
an essential singularity with respect to $E$. We discuss the asymptotic
evaluation in terms of the contour integral for arbitrary $t$ in
Sec.~\ref{subsec:Tunneling-contribution}.

We note that the difference of Berry connection $A_{++}-A_{--}$ that appears in Eq.~(\ref{eq:Wt}) and seems gauge dependent can be rewritten by a gauge invariant quantity, i.e., so called ``shift vector'',
\begin{equation}
R=A_{++}-A_{--}-\partial_{k}\arg A_{+-}.
\end{equation}
This allows us to rewrite $W(t)$ as~\cite{Kitamura2020}
\begin{equation}
W(t)=E|A_{+-}(k-Et)|e^{i\int_{t_0}^{t}dt^{\prime}(\varepsilon_{+}-\varepsilon_{-}+ER)+i\arg A_{+-}(0)}.\label{eq:Wt-R}
\end{equation}
This shift vector is known to appear in formulation of the second order nonlinear optical response called ``shift current''~\cite{Sipe,Young-Rappe,Morimoto-Nagaosa16}, and is a geometrical quantity that measures the real space shift between the centers of valence and conduction wavefunctions. As we show in Sec.~\ref{subsec:Nonreciprocal-transport}, shift vector also governs nonreciprocity in the tunneling current.

\subsection{Nonequilibrium Green's function\label{subsec:Nonequilibrium-Green's-function}}

Now we introduce a particle reservoir (so called B\"{u}ttiker bath~\cite{Buttiker86,Aoki2014}) and consider a nonequilibrium
steady state of the tunneling problem. We consider an open system
described by 
\begin{align}
\hat{\mathcal{H}}(t) & =\sum_{k}\hat{\mathcal{H}}_{k}(t),\label{eq:ham0}\\
\hat{\mathcal{H}}_{k}(t) & =\sum_{\sigma\sigma^{\prime}}\langle\sigma|H(k-Et)|\sigma^{\prime}\rangle\hat{c}_{k\sigma}^{\dagger}(t)\hat{c}_{k\sigma^{\prime}}(t)\nonumber \\
 & +\sum_{\sigma p}\omega_{p}\hat{b}_{k\sigma p}^{\dagger}(t)\hat{b}_{k\sigma p}(t)+\sum_{\sigma p}V_{p}\hat{b}_{k\sigma p}^{\dagger}(t)\hat{c}_{k\sigma}(t)+h.c.\label{eq:ham}
\end{align}
Here, $H(k)$ is the Hamiltonian Eq.~(\ref{eq:hamiltonian-isolated})
defined in the previous subsection, and $\sigma=\uparrow,\downarrow$
is the pseudospin spanning the Hilbert space of $2\times2$ Hamiltonian
$H(k)$ (corresponding to a sublattice, for instance). Note that we
neglect the real spin of the electron here for simplicity. $\hat{c}_{k\sigma}(t)$
annihilates an electron with momentum $k$ and pseudospin $\sigma,$
while $\hat{b}_{k\sigma p}(t)$ annihilates an electron in a fermionic
heat reservoir whose mode energy is $\omega_{p}$. The second-quantized
operators are represented in the Heisenberg representation, and denoted
by hats. The spectral density of the fermionic reservoir is assumed to satisfy the broadband condition,
\begin{equation}
\sum_{p}\pi|V_{p}|^{2}\delta(\omega-\omega_{p})=\Gamma=const,\label{eq:markov}
\end{equation}
which makes the dissipative dynamics of electrons Markovian [See Eq.~(\ref{eq:markov-trick})].

As we are interested in the tunneling process, it is natural to
introduce the snapshot basis as in the isolated cases. Namely, we
introduce an expansion of the field operator into the snapshot eigenstates
as 
\begin{align}
\hat{c}_{k\sigma}(t) & =\sum_{\alpha}\hat{\psi}_{\alpha,k}(t)\langle\sigma|\psi_{\alpha,k}(t)\rangle\label{eq:snapshot-expansion}\\
 & =\sum_{\alpha}\hat{\psi}_{\alpha,k}(t)\langle\sigma|u_{\alpha,k-Et}\rangle e^{-i\int_{t_0}^{t}dt^{\prime}(\varepsilon_{\alpha}+EA_{\alpha\alpha})},\\
\hat{\psi}_{\alpha,k}(t) & =\sum_{\sigma}\langle\psi_{\alpha,k}(t)|\sigma\rangle\hat{c}_{k\sigma}(t).
\end{align}

As the fermions in the reservoir are noninteracting, one can trace them out.
As a result, they are embedded in a self energy in terms of nonequilibrium Green's function.
By inserting the above transformation to the snapshot basis into the
well-known formula for the self energy (in the real-time representation
with the original basis), we obtain~\cite{Rammer86,Jauho94}
\begin{align}
G^{R}(t,t^{\prime}) & =G_{0}^{R}(t,t^{\prime})e^{-\Gamma(t-t^{\prime})},\label{eq:GR}\\
G^{A}(t,t^{\prime}) & =G_{0}^{A}(t,t^{\prime})e^{-\Gamma(t^{\prime}-t)},\label{eq:GA}\\
G^{<}(t,t^{\prime}) & =(G^R\ast\Sigma^{<}\ast G^A)(t,t^{\prime})\\&\coloneqq\int d\tau d\tau^{\prime}G^{R}(t,\tau)\Sigma^{<}(\tau,\tau^{\prime})G^{A}(\tau^{\prime},t^{\prime}),\label{eq:Glesser}
\end{align}
for the retarded, advanced, and lesser Green's function, which are
defined as $[G^{R}(t,t^{\prime})]_{\alpha\beta}=[G^{A}(t^{\prime},t)]_{\beta\alpha}^{\ast}=-i\langle\{\hat{\psi}_{\alpha,k}(t),\psi_{\beta,k}^{\dagger}(t^{\prime})\}\rangle\Theta(t-t^{\prime}),[G^{<}(t,t^{\prime})]_{\alpha\beta}=i\langle\psi_{\beta,k}^{\dagger}(t^{\prime})\hat{\psi}_{\alpha,k}(t)\rangle$ 
with $\Theta(t)=(1+\text{sgn}(t))/2$ being the step function.
 Here, $G_{0}^{R,A}$ denotes the Green's functions of the isolated
system. The lesser Green's function $G^{<}$ is a particularly interesting quantity as it describes the electron occupation in the nonequilibrium states. The lesser component of the self energy reads
\begin{align}
[\Sigma^{<}(\tau,\tau^{\prime})]_{\alpha\beta} & =i2\Gamma\int\dfrac{d\omega}{2\pi}e^{-i\omega(\tau-\tau^{\prime})}f_{D}(\omega)\langle\psi_{\alpha,k}(\tau)|\psi_{\beta,k}(\tau^{\prime})\rangle\label{eq:Sigma-lesser}
\end{align}
with $f_{D}$ being the Fermi-Dirac distribution function. We have
omitted the interval of integration $(-\infty,\infty)$ for the $\tau,\tau^{\prime},\omega$
integral. While this transformation is straightforward, we also provide
a derivation using the Heisenberg equation in Appendix~\ref{sec:Derivation-of-the}
for completeness.

To complete the framework, we need to specify the retarded Green's functions of the isolated
system $G_{0}^{R}(t,t^{\prime})$.
As $a_{\pm}(t)$ is the solution of the time evolution Eq.~(\ref{eq:eom-ap}),
one can explicitly construct the retarded Green's function of
the isolated system using a unitary matrix
\begin{equation}
U(t)=\begin{pmatrix}a_{-}^{\ast}(t) & a_{+}(t)\\
-a_{+}^{\ast}(t) & a_{-}(t)
\end{pmatrix},\label{eq:unitary}
\end{equation}
which satisfies 
\begin{align}
i\dot{U}(t) & =\begin{pmatrix}0 & W(t)\\
W^{\ast}(t) & 0
\end{pmatrix}U(t).
\end{align}
One can easily check that $G_{0}^{R}(t,t^{\prime})$ is represented
as 
\begin{equation}
G_{0}^{R}(t,t^{\prime})=-iU(t)U^{\dagger}(t^{\prime})\Theta(t-t^{\prime}).\label{eq:G0R}
\end{equation}
See also Appendix~\ref{sec:Derivation-of-the}. 

To summarize, the nonequilibrium Green's function of the open system
$G^{<}$ can be evaluated by, (i) computing the time evolution of
the isolated system Eq.~(\ref{eq:eom-ap}) to obtain $a_{\pm}$ and
construct $G_{0}^{R}$, and (ii) computing convolution of $\Sigma^{<}$
by performing $\tau,\tau^{\prime}$ and $\omega$ integrals in Eqs.~(\ref{eq:Glesser}),
(\ref{eq:Sigma-lesser}). We provide analytic expressions for the
outcome of this framework using various asymptotic methods in the
next section. 

Before closing the subsection, we remark that the nonequilibrium Green's
function is time dependent, nevertheless it represents a steady state.
This is because we focus on a single electron with a particular momentum $k$ (at
$t=0$), while the (steady) many-body state consists of electrons with various momenta.
In other words, the Green's function we consider here is that for $\hat{\mathcal{H}}_k(t)$,
while the physical system is given by $\hat{\mathcal{H}}(t)=\sum_k\hat{\mathcal{H}}_k(t)$ [See Eqs.~(\ref{eq:ham0}),~(\ref{eq:ham})].
Many-body expectation values, which are given as a momentum average of single-electron expectation values,
are indeed time independent since the direct relation between momentum and time, $k(t)=k-Et$,
makes momentum average identical to time average.

\subsection{Numerical calculation\label{subsec:Numerical-calculation}}

Here we use the above framework for performing numerical calculations,
and see the influence of the reservoir on the tunneling electrons.
We calculate the carrier density $n_+(t)$ as a transient occupation of
a single electron on the upper band,
\begin{equation}
n_+(t)=\langle\hat{\psi}_{+,k}^{\dagger}(t)\hat{\psi}_{+,k}(t)\rangle=\text{Im}[G^{<}(t,t)]_{++},
\end{equation}
which can be translated into the momentum distribution of the excited
electrons of the whole system. The carrier density $n_+(t)$ can be regarded as a counterpart of the (transient) tunneling probability in the case of isolated systems.

\begin{figure}
\centering{}\includegraphics[width=0.85\linewidth]{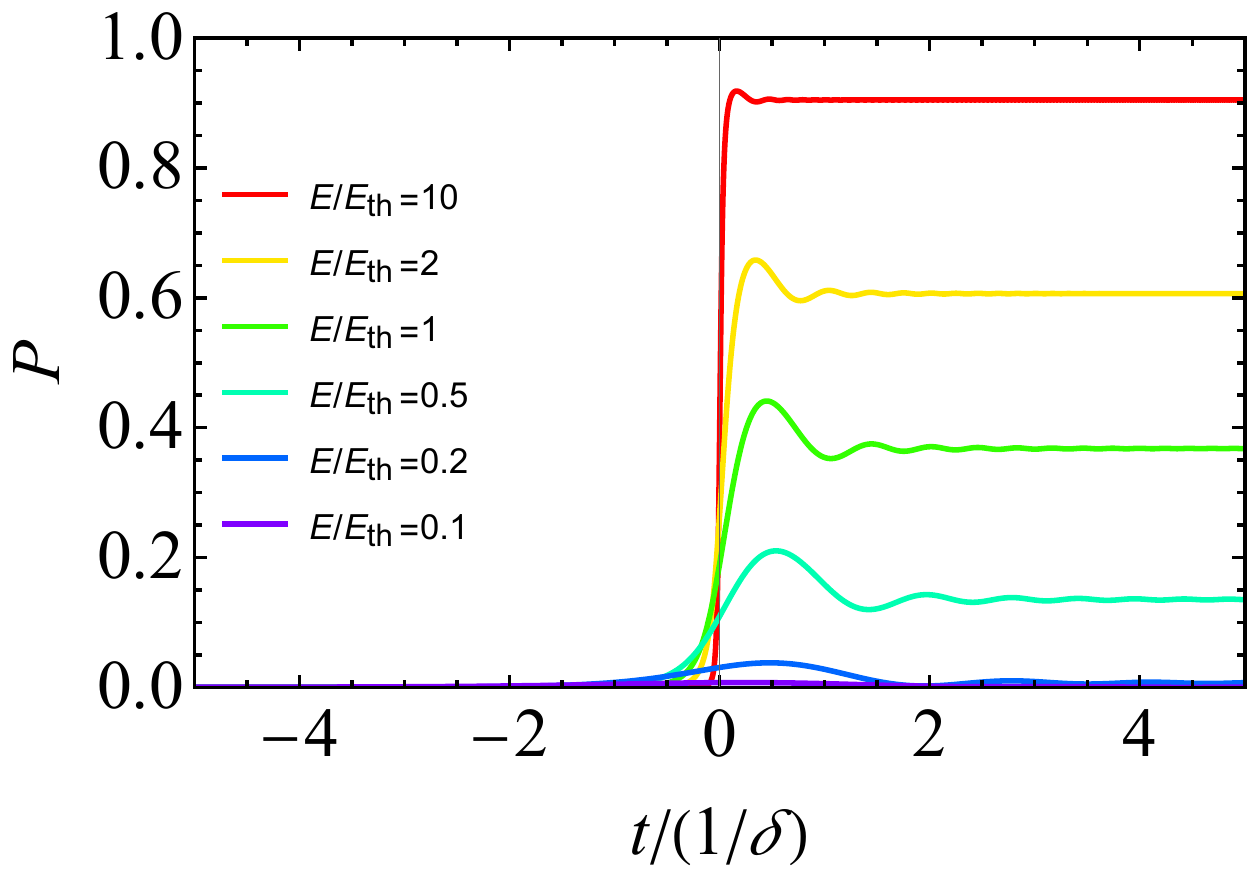}\caption{\label{fig:P-isolated}Tunneling probability $P(t)=|a_{+}(t)|^{2}$
of the isolated system as a function of time $t$, for the Landau-Zener
model. $E_{\text{th}}=\pi\delta^{2}/v$. }
\end{figure}

As a typical example, we consider the Landau-Zener model
\begin{equation}
H(k)=\begin{pmatrix}-vk & \delta\\
\delta & vk
\end{pmatrix},
\end{equation}
whose time evolution Eq.~(\ref{eq:eom-ap}) is known to be exactly-solvable~\cite{Zener32,DeGrandi2010}.  
Let us discuss the properties of the isolated case first. The tunneling
probability of the isolated case $P(t)=|a_{+}(t)|^{2}$ in the $t\rightarrow\infty$
limit is given as
\begin{equation}
P(t\rightarrow\infty)=e^{-E_{\text{th}}/E}=\exp\left(-\dfrac{\pi\delta^{2}}{vE}\right),
\end{equation}
which can also be exactly reproduced by the DDP method. The transient
dynamics is also important for characterizing the tunneling process.
We plot the tunneling probability $P(t)=|a_{+}(t)|^{2}$ as a function
of $t$ in Fig.~\ref{fig:P-isolated}, where we set $k(t=0)=0$.
It shows that the tunneling mainly occurs when the electron passes
through the gap minimum ($t=0$). In particular, the tunneling probability
approaches to the step function $\Theta(t)$ asymptotically in the
strong field limit. On the other hand, in the intermediate regime,
the tunneling probability undergoes an overshoot behavior within the
time scale of $\sim1/\sqrt{vE}$, before converging to the final value. 

\begin{figure}
\centering{}\includegraphics[width=0.85\linewidth]{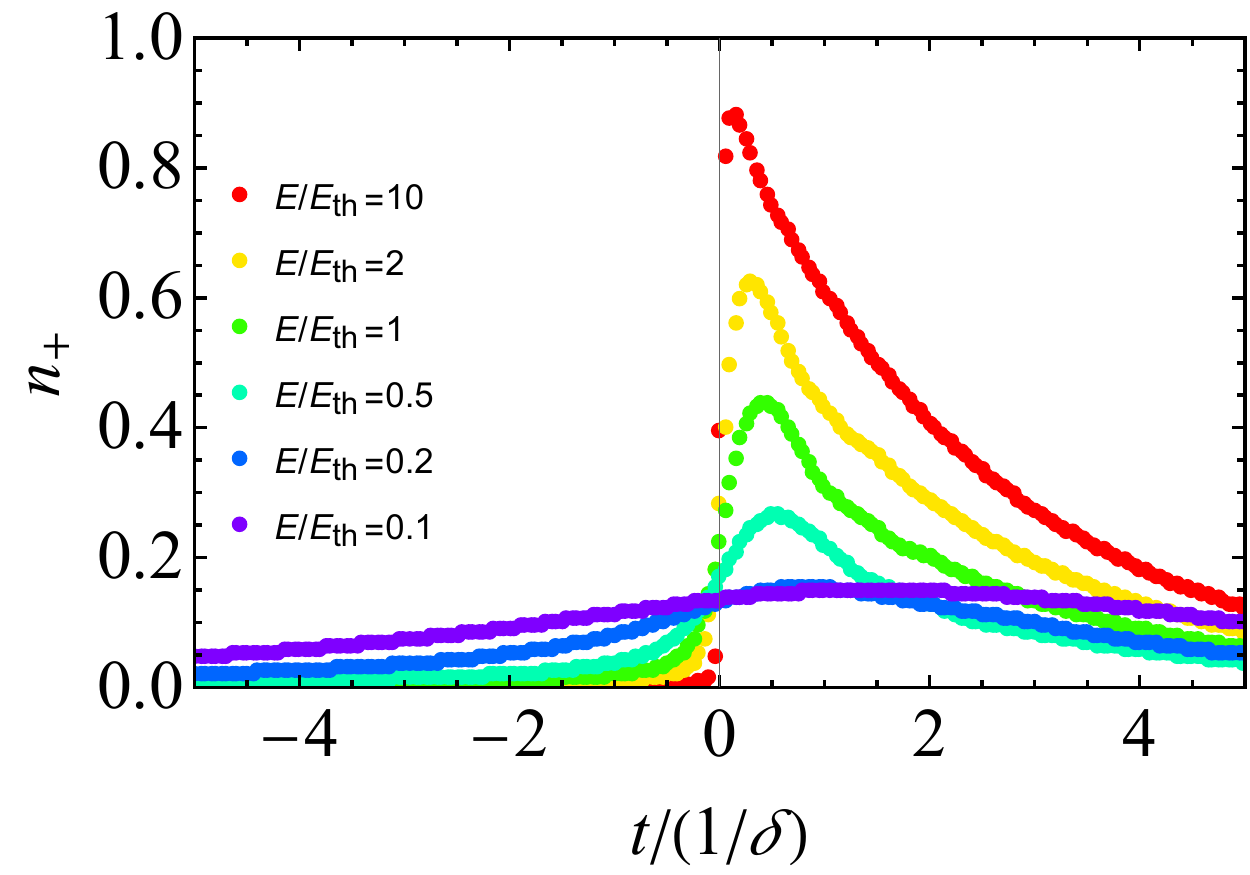}\caption{\label{fig:P-open}Carrier density $n_+(t)=\text{Im}[G^{<}(t,t)]_{++}$
of the open system as a function of time $t$, for the Landau-Zener
model. $E_{\text{th}}=\pi\delta^{2}/v$. } 
\end{figure}

Now, let us see how the carrier density (tunneling probability) is modified in the
presence of the fermionic reservoir. We plot the numerically-calculated transient
occupation of a single electron on the upper band in the open system $n_+(t)=\text{Im}[G^{<}(t,t)]_{++}$
in Fig.~\ref{fig:P-open}. Here, we set the temperature of the fermionic 
reservoir as $k_{B}T=0.5\delta$, which is relatively high, and $\Gamma=0.2\delta$.
We can find two qualitatively different regimes. One is the low-field
regime, where the tunneling amplitude in the isolated case is negligible
compared with the thermal excitation. In this regime, the system should
be well described by the perturbative treatment using the Boltzmann
equation, where the distribution of the electron follows the equilibrium
one with a drift of the momentum. On the other hand, as one increases
the field strength, the nonperturbative tunneling process becomes
dominant, and a jump in the carrier density evolves at $t=0$. This generated
carrier at the gap minimum gradually relaxes due to the coupling to
the fermionic reservoir [See also Fig.~\ref{fig:schematic}].

These features in the open system are expected to be universal in
generic gapped systems, and to be captured qualitatively by analytic
formulae using appropriate approximations, which we discuss in the
next section.

\section{Asymptotic evaluation\label{sec:Asymptotic-evaluation}}

\subsection{Overview\label{subsec:Overview}}

In this section, we evaluate the nonequilibrium Green's function Eq.~(\ref{eq:Glesser})
derived in the previous section, in an analytic manner using various
approximations. 
Let us begin with a brief overview of our derivation of Green's functions, before going into the details of calculations presented in the next subsections.
First, as a general remark, we note that the approximations
we adopt are mainly based on asymptotic expansions, as in the DDP method
in isolated systems. In contrast to usual Taylor series that has a finite convergence radius, these approximations are not necessarily improved by including the higher-order correction.
Thus we have to be careful on the condition when the approximation is justified.

We first consider the adiabatic limit and try to reproduce the low-field
regime. Since the dynamics of the isolated system is trivial there,
the central issue here is how to approximate $\tau,\tau^{\prime}$
and $\omega$ integrals in Eqs.~(\ref{eq:Glesser}), (\ref{eq:Sigma-lesser}).
As we are considering the adiabatic limit, where the time scale associated
with the change of the parameter is slow enough, we assume that it
is also slower than the decay time $\sim1/\Gamma$.
We can perform the $\tau,\tau^{\prime}$ integrals in a form of an
asymptotic series, which can be truncated in a low order if the above
assumption holds. This corresponds to the gradient expansion known
in the quantum kinetic theory~\cite{Rammer86}, which is employed for deriving the
quantum Boltzmann equation. Indeed, by performing $\omega$ integral
in terms of the residue integral, we obtain 
\begin{align}
\left[G_{\text{ad}}^{<}(t,t)\right]_{\pm\pm} & \simeq if_{D}(\varepsilon_{\pm}(t))+if_{D}^{\prime}(\varepsilon_{\pm}(t))\partial_{k}\varepsilon_{\pm}(t)\dfrac{E}{2\Gamma}
\end{align}
at the leading order, which coincides with the result of the Boltzmann
equation with the relaxation-time approximation. This is discussed
in Sec.~\ref{subsec:quasiequilibrium}. We also show that the above
approximation quantitatively deviates from the numerical result in
an insulating system due to the nonperturbative contribution.

Next we consider the tunneling contribution by extending the above
result. As we need to construct the Green's function $G_{0}^{R}$,
we have to calculate $a_{+}(t)$ at generic time $t$ as opposed to
the conventional tunneling problem where one considers only the $t\rightarrow\infty$
limit. According to the Lefschetz thimble approach recently proposed
for the tunneling problem~\cite{Fukushima2020}, the asymptotic form
for the nonperturbative component should be given as 
\begin{align}
a_{+}(t) & \simeq\sqrt{P_{0}}\Theta(t),
\end{align}
where $P_{0}$ is the tunneling probability of the isolated system
in the $t\rightarrow\infty$ limit. While
the discontinuity due to the step function is not present in the actual
solution, this approximates the rapid increase at $t=0$ that appeared
in Fig.~\ref{fig:P-isolated}. With this correction we can approximate
the nonequilibrium Green's function as 
\begin{multline}
G^{<}(t,t)\simeq G_{\text{ad}}^{<}(t,t)+i\begin{pmatrix}P_{0} & \sqrt{P_{0}}\\
\sqrt{P_{0}} & -P_{0}
\end{pmatrix}\\\times(f_{D}(\varepsilon_{-}(0))-f_{D}(\varepsilon_{+}(0)))e^{-2\Gamma t}\Theta(t),
\end{multline}
where the second term describes the decay of the tunnel electron seen in Fig.~\ref{fig:P-open}. This is the key result of the present study, which we discuss in Sec.~\ref{subsec:Tunneling-contribution}.

\subsection{Adiabatic limit\label{subsec:quasiequilibrium}}

Let us consider a situation where the electric field is so weak that
the nonperturbative contribution to the Green's function can be neglected.
We consider the adiabatic limit, $E\rightarrow0$, where the isolated
Green's function becomes trivial since $a_{+}(t)=0,$ $a_{-}(t)=1$,
and $U(t)=I_{2\times2}$. In this limit, the lesser Green's function
reads
\begin{align}
\left[G_{\text{ad}}^{<}(t,t^{\prime})\right]_{\alpha\beta} & =i2\Gamma\int\dfrac{d\omega}{2\pi}f_{D}(\omega)e^{-i\omega(t-t^{\prime})}\langle\mathcal{L}_{\omega}\psi_{\alpha}(t)|\mathcal{L}_{\omega}\psi_{\beta}(t^{\prime})\rangle,\label{eq:Glesser-ad}
\end{align}
where $\mathcal{L}_{\omega}$ represents the Laplace transform (from $\tau$ to $\Gamma+i\omega$), 
\begin{equation}
|\mathcal{L}_{\omega}\psi_{\alpha}(t)\rangle\coloneqq\int_{0}^{\infty}d\tau|\psi_{\alpha,k}(t-\tau)\rangle e^{-(\Gamma+i\omega)\tau}.\label{eq:laplace}
\end{equation}

In this subsection we try to construct an adiabatic perturbation expansion
of the nonequilibrium Green's function. This can be done when the
relaxation time $1/(2\Gamma)$ is sufficiently shorter than the typical
time scale of adiabatic parameter change ($\propto1/E$). In such
a case, the Laplace transform Eq.~(\ref{eq:laplace}) can be evaluated
in an asymptotic series form as follows.

A straightforward and elementary approach to obtain an asymptotic
expansion is successive uses of integration by parts based on the
relation 
\begin{equation}
e^{-(\Gamma+i\omega)\tau-i\int_{t_0}^{t-\tau}dt^{\prime}\varepsilon_{\alpha}}=-\dfrac{\partial_{\tau}(e^{-(\Gamma+i\omega)\tau-i\int_{t_0}^{t-\tau}dt^{\prime}\varepsilon_{\alpha}})}{\Gamma+i\omega-i\varepsilon_{\alpha}(t-\tau)},
\end{equation}
where we have introduced a short-hand notation $\varepsilon_{\alpha}(t-\tau)=\varepsilon_{\alpha}(k-E(t-\tau))$.
Instead, here we use a more systematic approach in the following.

Since the integrand decays in the time scale of $1/\Gamma$, one can
Taylor-expand the slowly-changing part of the integrand around $\tau=0$
and perform the termwise Laplace transform, which yields the asymptotic
series solution. However, as can be seen in the definition Eq.~(\ref{eq:snapshot-basis}),
the integrand $|\psi_{\alpha,k}(t-\tau)\rangle$ has two different
time scales. One is the adiabatic time scale appearing via $k(t)=k-Et$,
while another is the time dependence due to the dynamical phase factor
$-i\int_{t_0}^{t}dt^{\prime}\varepsilon_{\alpha}$. The latter should
be separately treated in performing the Taylor expansion (at least at the leading order). To this
end, we introduce the slow component at time $t$ as 
\begin{equation}
|\overline{\psi}_{\alpha,k}(t,\tau)\rangle=|\psi_{\alpha,k}(t-\tau)\rangle e^{-i\varepsilon_{\alpha}(t)\tau},\label{eq:slow-component}
\end{equation}
where the additional phase factor cancels the dynamical phase around
$\tau=0$. One can easily check that $\partial_{\tau}|\overline{\psi}_{\alpha,k}(t,\tau)\rangle=\mathcal{O}(E)$.
Now,
by expanding the slow component $|\overline{\psi}_{\alpha,k}(t,\tau)\rangle$,
we obtain 
\begin{align}
|\mathcal{L}_{\omega}\psi_{\alpha}(t)\rangle & =\int_{0}^{\infty}d\tau|\overline{\psi}_{\alpha,k}(t,\tau)\rangle e^{-(\Gamma+i\omega-i\varepsilon_{\alpha}(t))\tau}\\
 & =\sum_{n=0}^{\infty}\left.\dfrac{1}{n!}\dfrac{\partial^{n}}{\partial\tau^{n}}|\overline{\psi}_{\alpha,k}(t,\tau)\rangle\right|_{\tau=0}\int_{0}^{\infty}d\tau\tau^n e^{-(\Gamma+i\omega-i\varepsilon_{\alpha}(t))\tau}
\end{align}
\begin{align}
 & =\sum_{n=0}^{\infty}\left.\dfrac{\partial^{n}}{\partial\tau^{n}}\dfrac{|\overline{\psi}_{\alpha,k}(t,\tau)\rangle}{(\Gamma+i\omega-i\varepsilon_{\alpha}(t))^{n+1}}\right|_{\tau=0}\\
 & =\exp\left[-\dfrac{\partial}{\partial s}\dfrac{\partial}{\partial\tau}\right]\left.\dfrac{|\overline{\psi}_{\alpha,k}(t,\tau)\rangle}{s+i\omega}\right|_{s=\Gamma-i\varepsilon_{\alpha}(t),\tau=0}.\label{eq:gradient}
\end{align}
Equation~(\ref{eq:Glesser-ad}) then reads
\begin{widetext}
\begin{equation}
\left[G_{\text{ad}}^{<}(t,t^{\prime})\right]_{\alpha\beta}=i2\Gamma \left.\exp\left[-\frac{\partial}{\partial s}\frac{\partial}{\partial\tau}-\frac{\partial}{\partial s^{\prime}}\frac{\partial}{\partial\tau^{\prime}}\right]I(s,s^{\prime})\langle\overline{\psi}_{\alpha,k}(t,\tau)|\overline{\psi}_{\beta,k}(t^{\prime},\tau^{\prime})\rangle \right|_{s=\Gamma+i\varepsilon_{\alpha}(t),s^{\prime}=\Gamma-i\varepsilon_{\beta}(t^{\prime}),\tau=\tau^{\prime}=0},
\end{equation}
\end{widetext}
where
\begin{equation}
I(s,s^{\prime})=\int\dfrac{d\omega}{2\pi}\dfrac{f_{D}(\omega)e^{-i\omega(t-t^{\prime})}}{(s-i\omega)(s^{\prime}+i\omega)}.\label{eq:iss}
\end{equation}

Let us evaluate the $\omega$ integral $I(s,s^\prime)$. In this subsection,
let us focus on the case $t=t^{\prime}$. The integration can be
performed using the residue integral as
\begin{equation}
I(s,s^{\prime})=\dfrac{1}{s+s^{\prime}}f_{\Gamma}(\text{Im}s,-\text{Im}s^{\prime}),\label{eq:iss2}
\end{equation}
by using $\text{Re}s=\text{Re}s^{\prime}=\Gamma>0$. Here,
$f_{\Gamma}(\varepsilon_{1},\varepsilon_{2})$ is given by 
\begin{equation}
f_{\Gamma}(\varepsilon_{1},\varepsilon_{2})=\dfrac{1}{2}-\dfrac{1}{2\pi i}\left(\Psi\left(\dfrac{1}{2}+\dfrac{\Gamma+i\varepsilon_{1}}{2\pi k_{B}T}\right)-\Psi\left(\dfrac{1}{2}+\dfrac{\Gamma-i\varepsilon_{2}}{2\pi k_{B}T}\right)\right),
\end{equation}
with $\Psi$ being the digamma function,
which can be regarded as a ``modified distribution function'' reflecting the presence of the fermionic reservoir. We note that 
\begin{equation}
\text{Re}f_{\Gamma}(\varepsilon_{1},\varepsilon_{2})=\dfrac{1}{2}(f_{\Gamma}(\varepsilon_{1},\varepsilon_{1})+f_{\Gamma}(\varepsilon_{2},\varepsilon_{2})),
\end{equation}
and $f_{\Gamma}(\varepsilon,\varepsilon)\rightarrow f_{D}(\varepsilon)$
as $\Gamma/k_{B}T\rightarrow0$. Namely, the present bath behaves
as an ideal bath when $\Gamma\ll k_{B}T$. 

Having completed three integrations, we can obtain the expression
for the lesser Green's function by evaluating $\exp[-\partial_{s}\partial_{\tau}-\partial_{s^{\prime}}\partial_{\tau^{\prime}}]$.
While the $s$ derivative of Eq.~(\ref{eq:iss2}) consists of that of
the distribution $f_{\Gamma}$ and that of the denominator $(s+s^{\prime})^{-1}$,
the former should be smaller since it is higher order in $\Gamma/k_{B}T$.
Thus we truncate the former series at the first order: 
\begin{align}
e^{-\partial_{s}\partial_{\tau}-\partial_{s^{\prime}}\partial_{\tau^{\prime}}}I(s,s^{\prime})=&
e^{i\partial_{\varepsilon_1}\partial_{\tau}-i\partial_{\varepsilon_2}\partial_{\tau^{\prime}}}
f_{\Gamma}\times e^{-\partial_{s}(\partial_{\tau}+\partial_{\tau^{\prime}})}(s+s^{\prime})^{-1}\\
\simeq&[f_{\Gamma}+i(\partial_{\varepsilon_{1}}f_{\Gamma}\partial_{\tau}-\partial_{\varepsilon_{2}}f_{\Gamma}\partial_{\tau^{\prime}})]\nonumber\\
 & \qquad\qquad \times e^{-\partial_{s}(\partial_{\tau}+\partial_{\tau^{\prime}})}(s+s^{\prime})^{-1},
\end{align}
which leads to 
\begin{align}
\left[G_{\text{ad}}^{<}(t,t)\right]_{\alpha\beta} & \simeq if_{\Gamma}(\varepsilon_{\alpha}(t),\varepsilon_{\alpha}(t))\delta_{\alpha\beta}\nonumber \\
 & -2\Gamma(\partial_{\varepsilon_{\alpha}}+\partial_{\varepsilon_{\beta}})f_{\Gamma}(\varepsilon_{\alpha}(t),\varepsilon_{\beta}(t))\nonumber \\
 & \times e^{-\partial_{\Gamma}\partial_{\tau}/2}\left.\dfrac{\langle\overline{\psi}_{\alpha,k}(t,\tau)|i\partial_{\tau}|\overline{\psi}_{\beta,k}(t,\tau)\rangle}{\varepsilon_{\alpha}(t)-\varepsilon_{\beta}(t)-i2\Gamma}\right|_{\tau=0}.\label{eq:g-ad-partial}
\end{align}
The remaining $\tau$ derivative can be evaluated using 
\begin{align}
\langle\overline{\psi}_{\alpha,k}(t,\tau)|i\partial_{\tau}|\overline{\psi}_{\alpha,k}(t,\tau)\rangle & =\varepsilon_{\alpha}(t)-\varepsilon_{\alpha}(t-\tau),\\
\langle\overline{\psi}_{+,k}(t,\tau)|i\partial_{\tau}|\overline{\psi}_{-,k}(t,\tau)\rangle & =W(t-\tau)e^{i(\varepsilon_{+}(t)-\varepsilon_{-}(t))\tau},
\end{align}
which results in, for the diagonal part, 
\begin{align}
\left[G_{\text{ad}}^{<}(t,t)\right]_{\pm\pm} & \simeq if_{D}(\varepsilon_{\pm}(t))+if_{D}^{\prime}(\varepsilon_{\pm}(t))\partial_{k}\varepsilon_{\pm}(t)\dfrac{E}{2\Gamma}\label{eq:boltzmann}
\end{align}
at the leading order, which reproduces the well-known result of the
Boltzmann equation with the relaxation-time approximation. One can neglect the offdiagonal
part,
\begin{align}
\left[G_{\text{ad}}^{<}(t,t)\right]_{+-} & \simeq-\dfrac{2\Gamma(\partial_{\varepsilon_{+}}+\partial_{\varepsilon_{-}})f_{\Gamma}(\varepsilon_{+}(t),\varepsilon_{-}(t))}{\varepsilon_{+}(t)-\varepsilon_{-}(t)-i2\Gamma}W(t),
\end{align}
which can be shown to be cancelled with the perturbative correction to $U(t)$. 

\begin{figure}
\begin{centering}
\includegraphics[width=0.85\linewidth]{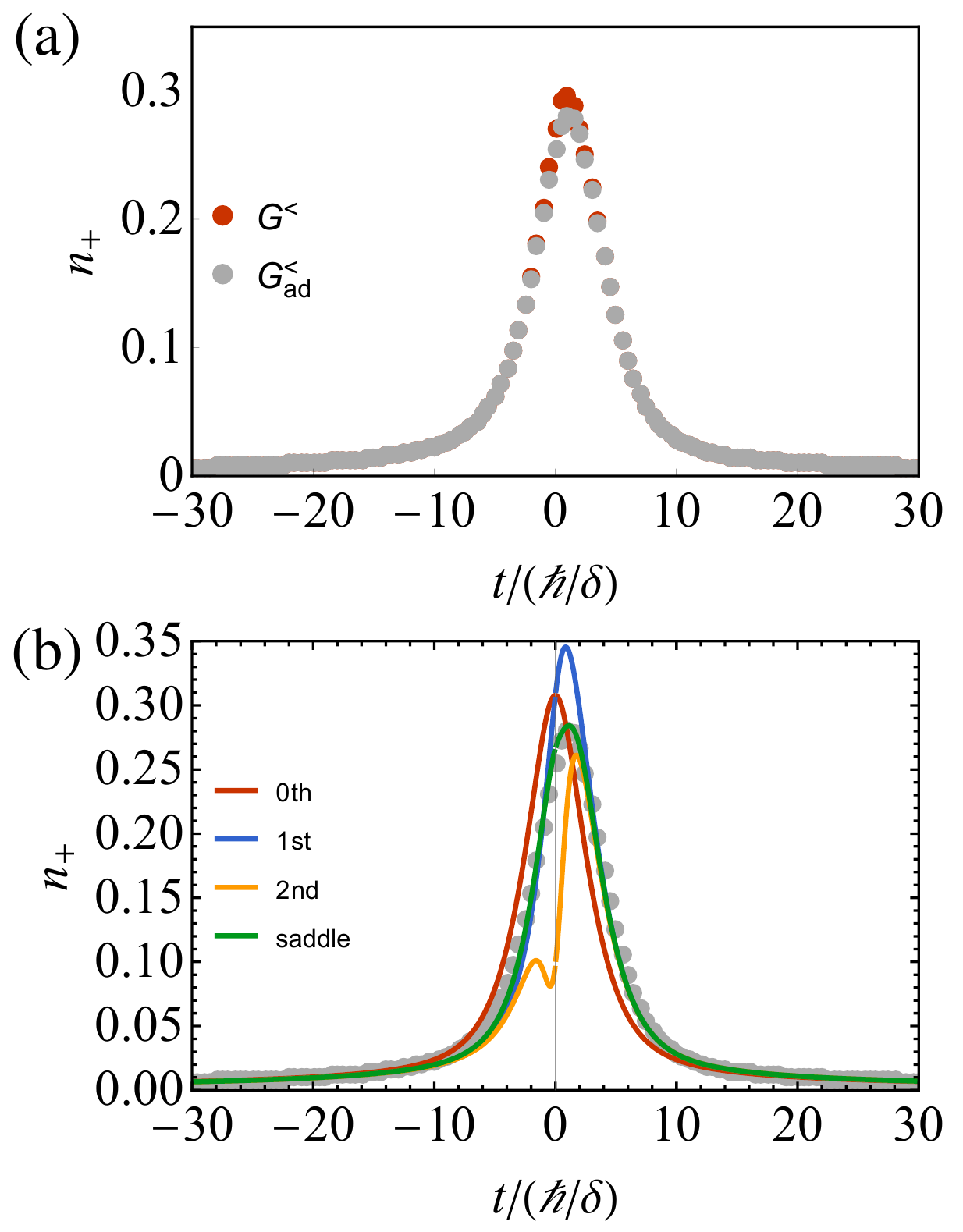}
\par\end{centering}
\centering{}\caption{\label{fig:P-adiabatic}Carrier density $n_+(t)=\text{Im}[G^{<}(t,t)]_{++}$
of the Landau-Zener model as a function of time. $\Gamma=0.4\delta$,
$k_{B}T=\delta$ and $E=0.2(\pi\delta^{2}/v)$. (a) Numerical calculation
based on the full Green's function Eq.~(\ref{eq:Glesser}) (red)
and based on the adiabatic component Eq.~(\ref{eq:Glesser-ad}) (gray).
(b) Comparison of the adiabatic component with the asymptotic expression
Eq.~(\ref{eq:g-ad-partial}). The Taylor expansion of $e^{-\partial_{\Gamma}\partial_{\tau}/2}$
is truncated at the $n$-th order. Green line is the result of the
saddle point approximation (See Appendix~\ref{sec:Nonperturbative-contribution-to}). }
\end{figure}

We examine the obtained formula by calculating the carrier 
density $n_+(t)=\text{Im}[G^{<}(t,t)]_{++}$ in Fig.~\ref{fig:P-adiabatic},
where we set $\Gamma=0.4\delta$, $k_{B}T=\delta$ and $E=0.2(\pi\delta^{2}/v)$
for the Landau-Zener model. As can be seen in the numerical result
plotted in Fig.~\ref{fig:P-adiabatic}(a), the results for the full expression of $G^{<}$,  Eq.~(\ref{eq:Glesser}), 
and $G_{\text{ad}}^{<}$ given by Eq.~(\ref{eq:Glesser-ad}) agree well, which implies that the thermal
excitation is the dominant mechanism for the carrier generation in
this parameter regime. We plot the result using the asymptotic expansion
Eq.~(\ref{eq:g-ad-partial}) truncated at zeroth, first and second
derivative with respect to $\tau$. The first-order formula reproduces
the numerical result semi-quantitatively. The second order correction
makes the result worse, which is characteristic to the asymptotic
expansion with vanishing convergent radius. One can also notice the
overestimation of the height of the peak. This deviation is related
to a nonperturbative effect peculiar to insulating systems, with which
the agreement is substantially improved as can be seen in the green
curve obtained with the saddle point method. We discuss details of
this effect in Appendix~\ref{sec:Nonperturbative-contribution-to}. 

\subsection{Tunneling contribution\label{subsec:Tunneling-contribution}}

As one decreases the temperature or increases the field strength,
the dominant mechanism for the carrier generation should switch from
the thermal excitation to the quantum tunneling, which is not taken
into account in the previous subsection. In this subsection, we consider
the nonperturbative tunneling contribution. Since the Green's function includes such nonperturbative contribution in the time evolution of the isolated system
$a_{\pm}(t)$, here we consider the first-order correction Eq.~(\ref{eq:ap1st})
in terms of the adiabatic perturbation. 

The central issue here is that we have to compute $a_{+}(t)$ as a
function of $t$, which is in contrast to the conventional tunneling
problem discussing the $t\rightarrow\infty$ limit. We first discuss
this using the Lefschetz thimble approach~\cite{Fukushima2020}.
Then we construct $G_{0}^{R}(t,t^{\prime})$ with the tunneling correction
and derive the formula for $G^{<}(t,t^{\prime})$. 

\subsubsection{thimble decomposition}

As is also known in the DDP method, it is essential to regard the
$t_{1}$ integral in Eq.~(\ref{eq:ap1st}) as a contour integral
of a complexified variable ($k-Et_{1}\rightarrow z_{1}$ here), in
capturing the nonperturbative nature of the tunneling probability.
The Lefschetz thimble method is a powerful tool in computing contour
integral, which provides a systematic decomposition of the contour
of integration $C_{0}$ (with $a_{+}(t)=\int_{C_{0}}dz_{1}e^{f(z_{1})}$)
into a deformed contour $C$ composed of the steepest descents of
$\text{Re}f(z_{1})$ that extend from saddle points (and the end point of $C_0$). See Refs.~\cite{Fukushima2020,Witten2010} and  Appendix~\ref{sec:Tunneling-amplitude-evaluated} for details. 
Because the steepest descent of $\text{Re}f(z_{1})$ coincides with the isopleth of $\text{Im}f(z_{1})$ due to the Cauchy-Riemann relations, the integrand along the deformed contour has no oscillation (as opposed to the original one) and is easier to evaluate.

\begin{figure}
\centering{}\includegraphics[width=0.75\linewidth]{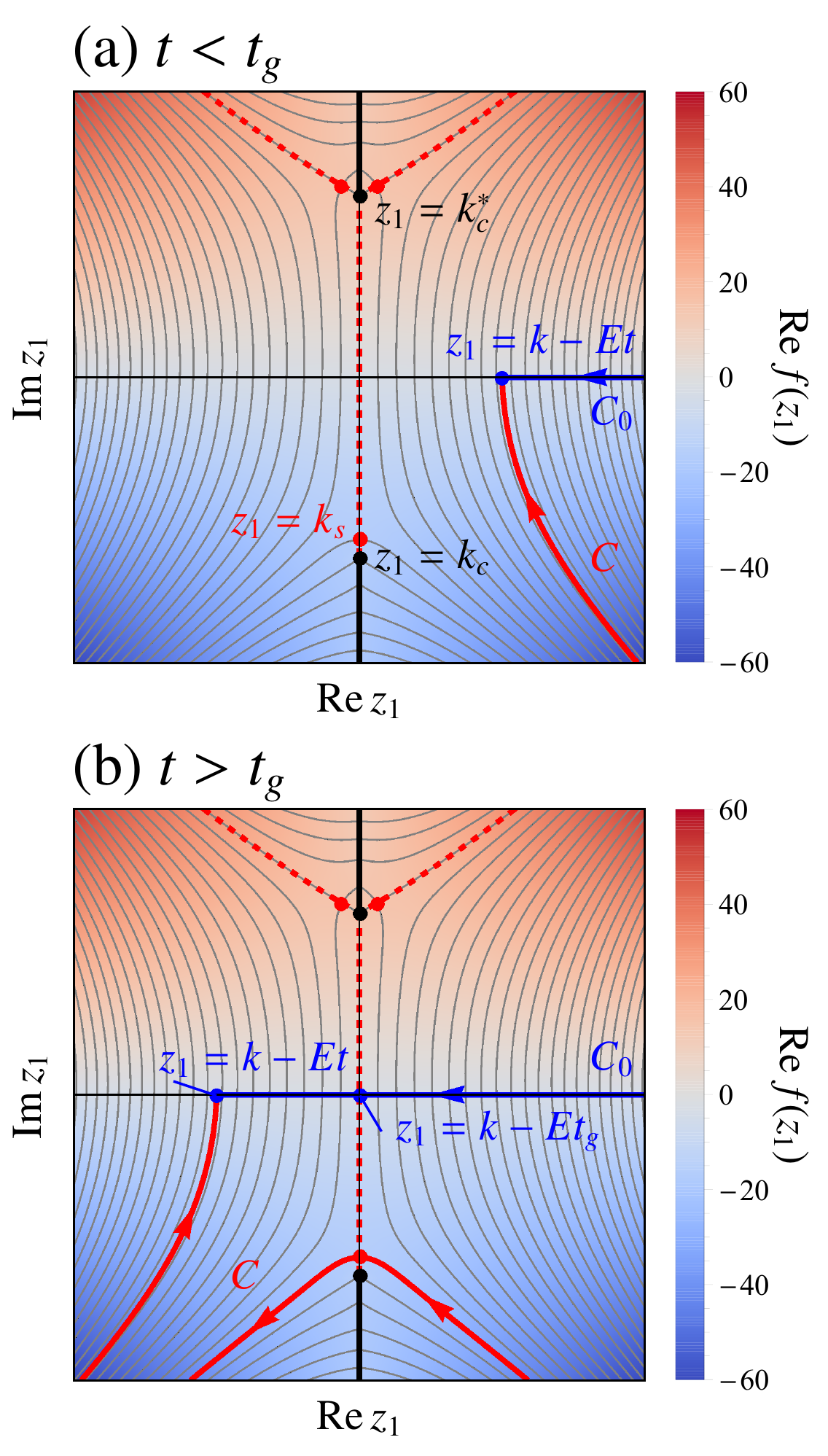}\caption{\label{fig:contour} $\text{Re}f(z_1)$ for $a_{+}(t)=\int_{C_{0}}dz_{1}e^{f(z_{1})}$ [See Eqs.~(\ref{eq:ap1st}), (\ref{eq:ap1st-cntr})], in the complexified momentum plane ($z_1=k-Et_1$). 
Gray lines are the steepest descents. The gap closing points $z_1=k_{c},k_c^\ast$ are indicated by
black dots, from which the branch cut drawn by the black thick
lines extends. The saddle points $z_1=k_s$ are marked with red dots, while red dashed lines are associated steepest ascents.
The original contour of integration $C_0$ indicated by blue line is deformed into $C$ composed of the steepest descents attached to saddle points (and the terminal point of the $C_0$), drawn as red solid lines. $t_g$ is defined via the crossing point $z_1=k-Et_g$ between the real axis and the red dashed line (steepest ascent). $t\gtrless t_g$ (whether $z_1=k-Et_g$ intersects with $C_0$ or not) determines whether the steepest descent attached to the saddle point ($z_1=k_s$) 
belongs to $C$ or not.}
\end{figure}

The saddle point is a special point where the steepest descent and
ascent join, whose position is obtained by solving $\partial_{z_{1}}f(z_{1})=0$.
In the present case, this equation reads 
\begin{equation}
\dfrac{\partial}{\partial z_{1}}\ln\tilde{A}_{+-}-i\dfrac{\Delta}{E}-iR=0,
\end{equation}
where $\tilde{A}_{+-}(z_{1})$ is the analytic continuation of the dipole matrix element $|A_{+-}(k-Et_{1})|$, $R=A_{++}-A_{--}-\partial_{k}\arg A_{+-}$ the shift vector, and $\Delta=\varepsilon_{+}-\varepsilon_{-}$.
As we show in Appendix~\ref{sec:Tunneling-amplitude-evaluated},
when $E$ is small enough, the solution $z_{1}=k_{s}$ can be found
in the vicinity of the gap closing point $z_{1}=k_{c}$ with $\Delta(k_{c})=0$
(i.e., where the second term vanishes). This can be seen in the plot of $\text{Re} f(z_1)$ for the Landau-Zener model ($E>0$), Fig.~\ref{fig:contour}, where the gap closing points and saddle points are marked with black and red points, respectively. 

According to the Lefschetz thimble method, the steepest descent attached
to a given saddle point belongs to the deformed contour $C$, if its
steepest ascent has an intersection with the original contour $C_{0}$,
as exemplified in Figs.~\ref{fig:contour}(a) and (b): The saddle
point in the lower half plane (marked with red dot) has a steepest
ascent parallel to the imaginary axis (red dashed line), which crosses
the real axis at $z=k-Et_{g}$ (the gap minimum point $\partial_{k}\Delta=0$,
represented by the blue dot in Fig.~\ref{fig:contour}(b)). As the original contour
$C_{0}$ (blue line) runs from $+\infty$ to $k-Et$, the steepest
descent has a contribution when $t>t_{g}$. Indeed the deformed contour
$C$ drawn by red curves is composed of two pieces in Fig.~\ref{fig:contour}(b)
with $t>t_{g}$, in contrast to (a) with $t<t_{g}$.
While there are also two saddle points in the upper half plane (and more on another Riemann surface),
they always have no contribution as their steepest ascents do not intersect with the real axis.

The saddle point contribution present in $t>t_{g}$ can be evaluated
approximately using Laplace's method, which results in 
\begin{align}
a_{+}(t) & \simeq\sqrt{P_{0}}\Theta(t)=e^{\text{Im}\int_{0}^{k_{c}}dk(\Delta/|E|+\text{sgn}(E)R)}\Theta(t).\label{eq:ap1st-ddp}
\end{align}
Here, for simplicity, we have set $t_{g}=0$ by shifting the origin of time, and 
set $\arg A_{+-}(k=0)$ such that the tunneling amplitude becomes real [See Eq.~(\ref{eq:ap1st-ddp-detail})].
We keep only the leading order in $E$ for the prefactor. 
See Appendix~\ref{sec:Tunneling-amplitude-evaluated} for details.

The discontinuous behavior $\Theta(t)$ roughly approximates the time
profile shown in Fig.~\ref{fig:P-isolated} if we neglect the overshoot
behavior in $0<t\lesssim1/\sqrt{vE}$. The overshoot behavior is related
to the last segment of the deformed contour $C$ (steepest ascent
toward the terminal point $z_{1}=k-Et$), although we neglect it in
this study. 
When $E$ is small enough, perturbative evaluation of the last segment yields
an $\mathcal{O}(E\sqrt{P_{0}})$ term to the Green's function, which reproduces the overshoot behavior, 
although it cannot capture the suppression in the strong $E$ regime.
We note that its contribution to the electric current is higher-order than the interband component (we derive below) with $\mathcal{O}(\sqrt{P_{0}})$.

\subsubsection{Green's function}

Let us evaluate the influence of the tunneling contribution Eq.~(\ref{eq:ap1st-ddp})
on the nonequilibrium Green's function. With this contribution, the
retarded Green's function of the isolated system reads 
\begin{equation}
G_{0}^{R}(t,t^{\prime})\simeq-i\Theta(t-t^{\prime})I_{2\times2}+i\begin{pmatrix}P_{0}/2 & -\sqrt{P_{0}}\\
\sqrt{P_{0}} & P_{0}/2
\end{pmatrix}\Theta(t)\Theta(-t^{\prime}),
\end{equation}
where the diagonal entries in the second term arise from the correction
to $a_{-}(t)$ that keeps the norm conservation $|a_{+}|^{2}+|a_{-}|^{2}=1$
up to $\mathcal{O}(P_{0})$.

Since the tunneling process is approximated to be instantaneous and
represented by the step function within the present approximation,
the second term can be rewritten in terms of $G_{0,\text{ad}}^{R}(0,t^{\prime})=-i\Theta(0-t^{\prime})I_{2\times2}$.
Then $G^{R}(t,t^{\prime})=G_{0}^{R}(t,t^{\prime})e^{-\Gamma(t-t^{\prime})}$
reads 
\begin{align}
G^{R}(t,t^{\prime})&= G_{\text{ad}}^{R}(t,t^{\prime})+M(t)\Theta(t)G_{\text{ad}}^{R}(0,t^{\prime})
\end{align}
where
\begin{align}
M(t)&=-\begin{pmatrix}P_{0}/2 & -\sqrt{P_{0}}\\
\sqrt{P_{0}} & P_{0}/2
\end{pmatrix}e^{-\Gamma t}
\end{align}
and  $G^{R}_{\text{ad}}(t,t^{\prime})=G_{0,\text{ad}}^{R}(t,t^{\prime})e^{-\Gamma(t-t^{\prime})}$.

By substituting this and $G^{A}(t,t^{\prime})=[G^{R}(t^{\prime},t)]^\dagger$
into Eq.~(\ref{eq:Glesser}), $G^{<}=G^{R}\ast\Sigma^{<}\ast G^{A}$,
we obtain $G^{<}$ with the tunneling correction, in terms of
 $G_{\text{ad}}^{<}=G_{\text{ad}}^{R}\ast\Sigma^{<}\ast G_{\text{ad}}^{A}$
(Here, $\ast$ denotes convolution in time and matrix product in the band index).
Namely, we can summarize the (equal-time) expression into 
\begin{equation}
G^{<}(t,t)=G_{\text{ad}}^{<}(t,t)+G_{\text{LZ}}^{<}(t,t)\Theta(t)
\end{equation}
with
\begin{align}
G_{\text{LZ}}^{<}(t,t)&=M(t)G_{\text{ad}}^{<}(0,0)M^{\dagger}(t)\nonumber\\&+M(t)G_{\text{ad}}^{<}(0,t)+G_{\text{ad}}^{<}(t,0)M^{\dagger}(t).
\end{align}
In particular, the diagonal component of the correction term $G_{\text{LZ}}^{<}(t,t)$
reads 
\begin{align}
\left[G_{\text{LZ}}^{<}(t,t)\right]_{\pm\pm} & =\left[G_{\text{ad}}^{<}(0,0)\right]_{\mp\mp}P_{0}e^{-2\Gamma t}\nonumber \\
 & -i\text{Im}\left[G_{\text{ad}}^{<}(t,0)\right]_{\pm\pm}P_{0}e^{-\Gamma t}\nonumber \\
 & \pm2i\text{Im}\left[G_{\text{ad}}^{<}(t,0)\right]_{\pm\mp}\sqrt{P_{0}}e^{-\Gamma t}.
\end{align}

Since we have evaluated the equal-time expression $G_{\text{ad}}^{<}(t,t)$
in the previous subsection, we have to evaluate the adiabatic Green's
function $G_{\text{ad}}^{<}(t,t^{\prime})$ with $t>t^{\prime}=0$ here.
If we evaluate Eq.~(\ref{eq:iss}) with $t>t^{\prime}$, we obtain
\begin{align}
I(s,s^{\prime}) & =\dfrac{f_{D}(-is)e^{-s(t-t^{\prime})}}{s+s^{\prime}}+\sum_{n=0}^{\infty}\dfrac{ik_{B}Te^{-\omega_{n}(t-t^{\prime})}}{(s-\omega_{n})(s^{\prime}+\omega_{n})},
\end{align}
where $\omega_{n}=(2n+1)\pi k_{B}T$ is the Matsubara frequency. Since
we are considering $k_{B}T\gg\Gamma=\text{Re}s$, the second term
is negligible for $t-t^{\prime}\neq0$. In addition to the $s$ derivative
of the distribution $f_{D}$ and the denominator $(s+s^{\prime})^{-1}$, 
we have that of $e^{-s(t-t^{\prime})}$ in the evaluation of  $e^{-\partial_{s}\partial_{\tau}-\partial_{s^{\prime}}\partial_{\tau^{\prime}}}I(s,s^{\prime})$ in the present case. 
This contribution is problematic
when $t-t^{\prime}$ is large, since 
\begin{equation}
e^{-\partial_{s}\partial_{\tau}}e^{-s(t-t^{\prime})}=e^{-s(t-t^{\prime})}e^{-\partial_{s}\partial_{\tau}}e^{(t-t^{\prime})\partial_{\tau}}
\end{equation}
acts as a time-translation operator for $\tau$. This leads to the
breakdown of the assumption that $\tau$ is small, which is necessary
for performing the gradient expansion Eq.~(\ref{eq:gradient}). To
cancel this time translation effect, we need to choose the slow component
as
\begin{equation}
|\psi_{\alpha,k}(t-\tau)\rangle=|\overline{\psi}_{\alpha,k}(t^{\prime},\tau-(t-t^{\prime}))\rangle e^{i\varepsilon_{\alpha}(t^{\prime})(\tau-(t-t^{\prime}))}.
\end{equation}
For details see Appendix~\ref{sec:time-difference}. 
Then, as the remaining factors
in $I(s,s^{\prime})$ are the same as in the previous calculation,
we arrive at a similar expression as Eq.~(\ref{eq:g-ad-partial}), 
\begin{align}
\left[G_{\text{ad}}^{<}(t,t^{\prime})\right]_{\alpha\beta} & \simeq if_{D}(\varepsilon_{\alpha}(t^{\prime})-i\Gamma)e^{-\Gamma(t-t^{\prime})}\delta_{\alpha\beta}\nonumber \\
 & -2\Gamma f_{D}^{\prime}(\varepsilon_{\alpha}(t^{\prime})-i\Gamma)e^{-\Gamma(t-t^{\prime})}\nonumber \\
 & \times e^{-\partial_{\Gamma}\partial_{\tau}/2}\left.\dfrac{\langle\overline{\psi}_{\alpha,k}(t^{\prime},\tau)|i\partial_{\tau}|\overline{\psi}_{\beta,k}(t^{\prime},\tau)\rangle}{\varepsilon_{\alpha}(t^{\prime})-\varepsilon_{\beta}(t^{\prime})-2i\Gamma}\right|_{\tau=0}.
\end{align}
As the drift correction $\propto f_{D}^{\prime}$ is less relevant
when $E$ is increased (correction may make the asymptotic expansion
worse), let us consider only the first term. The correction to the
nonequilibrium Green's function reads
\begin{equation}
\left[G_{\text{LZ}}^{<}(t,t)\right]_{\pm\pm}\Theta(t)=\pm i(f_{D}(\varepsilon_{-}(0))-f_{D}(\varepsilon_{+}(0)))P_{0}e^{-2\Gamma t}\Theta(t).
\end{equation}
The physical meaning of this expression is apparent. The tunneling
occurs at $t=0$ with probability $P_{0}$, which is instantaneous
and governed by the quasi-equilibrium distribution at $t=0$ (although
this is an approximation). This contribution decays in the time scale
of $1/(2\Gamma)$, as the excited electrons are relaxed due to the dissipation to the heat bath. 
This picture is schematically summarized in Fig.~\ref{fig:schematic}.

In the same way, one can calculate the offdiagonal part as 
\begin{align}
\left[G_{\text{LZ}}^{<}(t,t)\right]_{+-} & \simeq\left(\left[G_{\text{ad}}^{<}(0,t)\right]_{--}-\left[G_{\text{ad}}^{<}(t,0)\right]_{++}\right)\sqrt{P_{0}}e^{-\Gamma t}\\
 & \simeq i(f_{D}(\varepsilon_{-}(0))-f_{D}(\varepsilon_{+}(0)))\sqrt{P_{0}}e^{-2\Gamma t},
\end{align}
where we have dropped $\mathcal{O}(EP_{0})$. It is worth noting that
the offdiagonal component has a halved nonperturbative exponent,
which implies that the interband current may be crucial for the transport
property. We compare intraband and interband contributions for the electric
current in Sec.~\ref{subsec:nonperturbative-electric}.

\section{Applications\label{sec:Applications}}

\subsection{Nonperturbative electric transport in band insulators\label{subsec:nonperturbative-electric}}

We have derived a formula for the nonequilibrium Green's function
with the nonperturbative correction in the previous section. The original
motivation to calculate this is to obtain the nonequilibrium distribution
of the electron and calculate physical observables, such as the electric
current. Here, let us evaluate the nonperturbative electric current
of the band insulators as an application of the present framework.
The velocity operator in the snapshot basis is expressed as 
\begin{align}
\hat{v} & =\sum_{\sigma\sigma^{\prime}}\langle\sigma|\partial_{k}H(k-Et)|\sigma^{\prime}\rangle\hat{c}_{k\sigma}^{\dagger}(t)\hat{c}_{k\sigma^{\prime}}(t)\\
 & =\begin{pmatrix}\hat{\psi}_{+,k}(t)\\
\hat{\psi}_{-,k}(t)
\end{pmatrix}^{\dagger}v(k-Et)\begin{pmatrix}\hat{\psi}_{+,k}(t)\\
\hat{\psi}_{-,k}(t)
\end{pmatrix}\\
 & \coloneqq \begin{pmatrix}\hat{\psi}_{+,k}(t)\\
\hat{\psi}_{-,k}(t)
\end{pmatrix}^{\dagger}\begin{pmatrix}\partial_{k}\varepsilon_{+} & i\Delta W/E\\
-i\Delta W^{\ast}/E & \partial_{k}\varepsilon_{-}
\end{pmatrix}\begin{pmatrix}\hat{\psi}_{+,k}(t)\\
\hat{\psi}_{-,k}(t)
\end{pmatrix},
\end{align}
where $\Delta=\varepsilon_{+}-\varepsilon_{-}$. Note that this expression
is exact for an arbitrary $E$ (i.e. it contains all the nonlinear
terms w.r.t the vector potential). 
We also note that $\arg W(t)$ depends on $\arg A_{+-}(k=0)$, which has been fixed such that the asymptotic form of $a_+(t)$ becomes real [See Eq.~(\ref{eq:Wt-mod})].
As we have mentioned in the end of Sec.~\ref{subsec:Nonequilibrium-Green's-function}, physical observables are given as a momentum average of the single-electron expectation value calculated with the nonequilibrium Green's function (and are thus time-independent). 
In the adiabatic limit, the electric
current is given as 
\begin{align}
J_{\text{ad}} & =-i\int\dfrac{dk}{2\pi}\text{Tr}[-vG_{\text{ad}}^{<}]\\
 & =-\dfrac{E}{2\Gamma}\int\dfrac{dk}{2\pi}\sum_{\alpha=\pm}(\partial_{k}\varepsilon_{\alpha})^{2}f_{D}^{\prime}(\varepsilon_{\alpha})\nonumber \\
 & -E\int\dfrac{dk}{2\pi}\dfrac{2\Gamma\Delta^{2}|A_{+-}|^{2}}{\Delta^{2}+4\Gamma^{2}}\sum_{\alpha=\pm}f_{D}^{\prime}(\varepsilon_{\alpha}),
\end{align}
which vanishes in the insulating system at the low temperature, as
$f_{D}^{\prime}$ becomes zero.  On the other hand, the nonperturbative
correction has a temperature dependence as 
\begin{equation}
J=(J_{\text{LZ}}^{(1)}+J_{\text{LZ}}^{(2)})(f_{D}(\varepsilon_{-}(0))-f_{D}(\varepsilon_{+}(0))),
\end{equation}
Here, the zero-temperature expressions $J_{\text{LZ}}^{(1)},J_{\text{LZ}}^{(2)}$
are the intraband and interband currents given as 
\begin{align}
J_{\text{LZ}}^{(1)} & =\mp P_{0}\int_{\mp\Lambda}^{0}\dfrac{dk}{2\pi}\partial_{k}\Delta e^{2\Gamma k/E},\label{eq:j1}\\
J_{\text{LZ}}^{(2)} & =2\sqrt{P_{0}}\text{Re}\int_{\mp\Lambda}^{0}\dfrac{dk}{2\pi}|A_{+-}|\Delta e^{-i\text{Re}\int_{k_c}^{k}dk^{\prime}(\Delta/E+R)+2\Gamma k/E}\label{eq:j2}
\end{align}
where $\pm=\text{sgn}(E)$~\footnote{While the present formula is justified 
only if the decay time $1/2\Gamma$ is shorter than the adiabatic time scale, 
its $\Gamma\rightarrow0$ limit partially reproduces the result in the ballistic limit as follows. 
The $\Gamma\rightarrow0$ expression for the intraband current is given as 
$J_{\text{LZ}}^{(1)} \rightarrow P_{0}(\Delta(\Lambda)-\Delta(0))/2\pi$. 
Here $\Delta(\Lambda)$  
can be regarded as the applied voltage in the Landauer picture, when two leads are sandwiching the system and the cutoff momentum is determined by the chemical potential of the leads. 
In this picture, we obtain the electric conductance as $P_0 \,(e^2/h)$, 
if we drop the small contribution from $\Delta(0)$. 
Also, when $\Gamma\rightarrow0$, the integral in the interband current Eq.~(\ref{eq:j2}) has 
no perturbative expression w.r.t. $E$. 
One can show by the saddle point method that $J_{\text{LZ}}^{(2)}$ is higher-order 
than $J_{\text{LZ}}^{(1)}$ in this limit. }.
Here, $\Lambda$ is a cutoff momentum, which should be replaced by $2\pi$ divided by the lattice constant in the case of lattice systems [See Sec.~\ref{subsec:Extension-to-lattice-system}]. $J_{\text{LZ}}^{(1)}$ is asymptotically evaluated as

\begin{equation}
J_{\text{LZ}}^{(1)}\sim\pm\dfrac{P_{0}}{2\pi}\left[-\dfrac{E}{2\Gamma}\dfrac{\partial\Delta}{\partial k}+\dfrac{E^{2}}{4\Gamma^{2}}\dfrac{\partial^{2}\Delta}{\partial k^{2}}-\dots\right]_{k=0}
\end{equation}
which survives since $f_{D}(\varepsilon_{-}(0))-f_{D}(\varepsilon_{+}(0))\sim1$. 
When the first derivative of $\Delta$ vanishes as in the Landau-Zener
model, the intraband tunneling current turns out to be proportional
to $E^{2}P_{0}$. One can evaluate the interband current $J_{\text{LZ}}^{(2)}$
by the similar asymptotic series expansion. The leading-order term
reads
\begin{align}
J_{\text{LZ}}^{(2)} & \sim\dfrac{\sqrt{P_{0}}}{\pi}\left[\dfrac{E|A_{+-}|2\Gamma\Delta}{\Delta^{2}+4\Gamma^{2}}\right]_{k=0},
\end{align}
where we have assumed $\text{Re}\int_{0}^{k_c}dk^{\prime}(\Delta/E+R)=0$ for simplicity.
While $J_{\text{LZ}}^{(2)}$ has a smaller power $E\sqrt{P_{0}}$
compared with $J_{\text{LZ}}^{(1)}$, the Lorentz factor makes the
value small when $\Gamma\ll\Delta$. Thus, whether the intraband or interband
effect is dominant depends on the strength of the dissipation. 

\begin{figure}
\centering{}\includegraphics[width=1\linewidth]{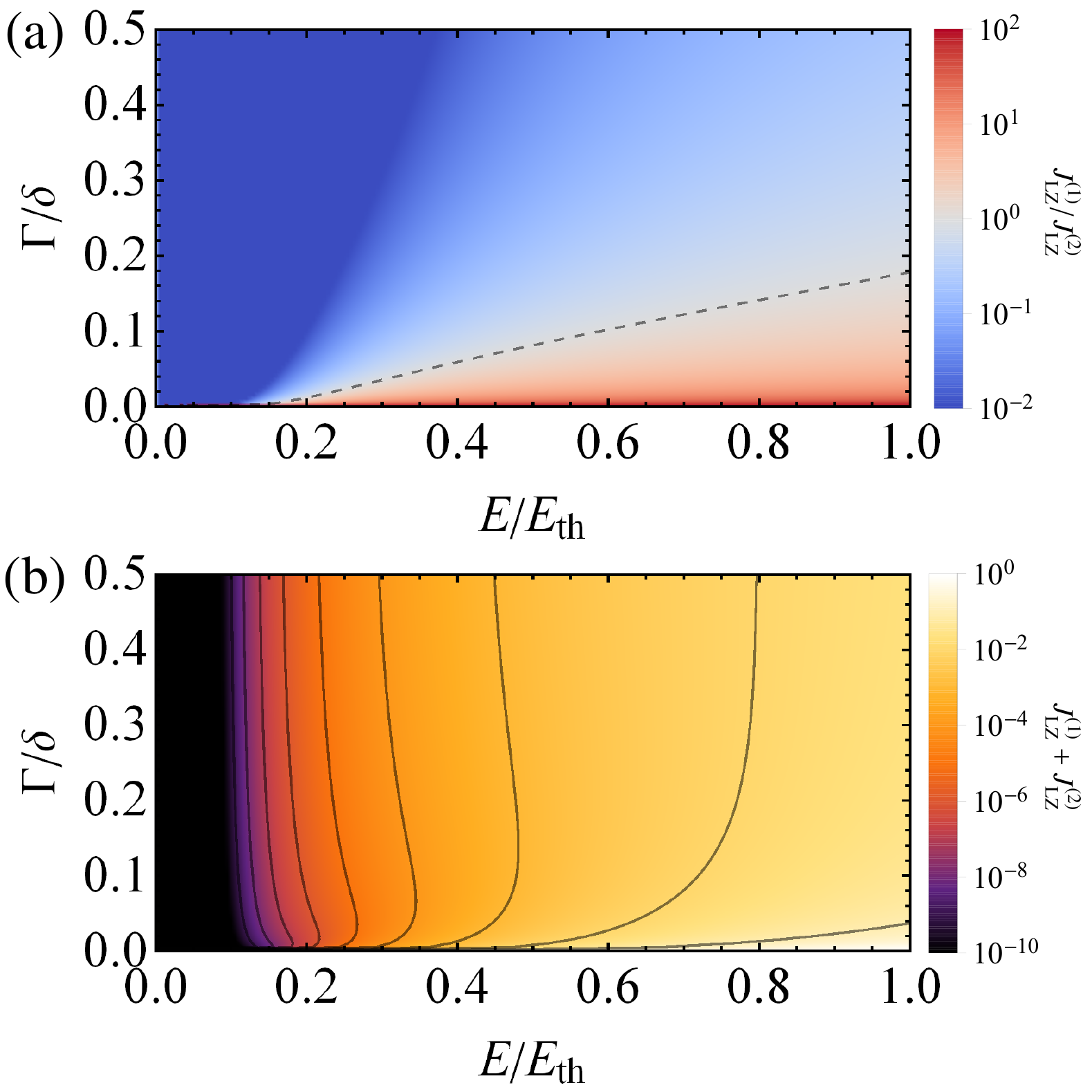}\caption{\label{fig:current} Electric current response of the nonequilibrium steady state for the Landau-Zener model attached to a  fermionic reservoir. (a) Ratio of the intraband and interband currents $J_{\text{LZ}}^{(1)}/J_{\text{LZ}}^{(2)}$
as a function of the electric field $E$ and the dissipation $\Gamma$. Dashed line indicates $J_{\text{LZ}}^{(1)}=J_{\text{LZ}}^{(2)}$. (b) 
$J_{\text{LZ}}^{(1)}+J_{\text{LZ}}^{(2)}$ as a function of the electric
field $E$ and the dissipation $\Gamma$.}
\end{figure}

We plot $J_{\text{LZ}}^{(1)}/J_{\text{LZ}}^{(2)}$ and $J_{\text{LZ}}^{(1)}+J_{\text{LZ}}^{(2)}$
for the Landau-Zener model in Figs.~\ref{fig:current}(a) and (b),
respectively, as functions of $E$ and $\Gamma$. Here, we have numerically
integrated Eqs.~(\ref{eq:j1}) and (\ref{eq:j2}). We find
that the interband current is dominant in a wide region of the parameter
space. The intraband current is dominant only when $\Gamma\lesssim0.1\delta$,
where one has a crossover from the interband-dominant to intraband-dominant
regime as increasing the field strength.

Such dominance of interband contribution to the current response cannot be captured by conventional analyses of tunneling processes that only focus on tunneling probability. 
Namely, the intraband contribution to the current can be deduced from the tunneling probability and group velocity. In contrast, the interband contribution, which turns out to be dominant in a wide parameter range, requires analysis of phase coherence of tunneling electrons, and cannot be captured only by looking at the tunneling probability. 
Thus our Green's function approach has an advantage in describing tunneling current response with an ability to incorporate the intraband and interband contributions on an equal footing.

\subsection{Nonreciprocal transport\label{subsec:Nonreciprocal-transport}}
\subsubsection{Nonreciprocal charge transport}
As we have revealed in the previous study~\cite{Kitamura2020}, the tunneling
probability $P_{0}$ has a geometric factor that involves the shift
vector $R$. In particular, for noncentrosymmetric systems, this factor
exhibits nonreciprocity (depends on the sign of $E$):
\begin{align}
\gamma_P\coloneqq&\frac{P_{0}(+|E|)}{P_{0}(-|E|)}=\frac{e^{2\text{Im}\int_{0}^{k_{c}}dk(\Delta/|E|+R)}}{e^{2\text{Im}\int_{0}^{k_{c}}dk(\Delta/|E|-R)}}=\exp\left[2\text{Im}\int_{k_c^\ast}^{k_{c}}dkR\right].
\end{align}
The shift vector $R$ is an odd function of $k$ when the system is
inversion-symmetric, and does not lead to nonreciprocity. In contrast, noncentrosymmetric
systems can host nonreciprocity arising from the geometric factor.

When the tunneling process is the main mechanism to generate carriers,
the nonreciprocity ratio $\gamma=J(+E)/J(-E)$ for the electric current should also be characterized
by that for tunneling probability $\gamma_P$. However, since the intraband and interband currents ($J_{\text{LZ}}^{(1)}$
and $J_{\text{LZ}}^{(2)}$ in the previous section) are respectively
proportional to $P_{0}$ and $\sqrt{P_{0}}$, the nonreciprocity ratio $\gamma$
for the electric current should undergo a crossover from $\sqrt{\gamma_P}$
to $\gamma_P$ when the dominant contribution is switched from
the interband to intraband current, e.g. by sweeping the strength
of the field~\footnote{
In a paper by two of the present authors [T.~Morimoto and N.~Nagaosa, ``Nonreciprocal current from electron interactions in noncentrosymmetric crystals: roles of time reversal symmetry and dissipation,'' \href{\doibase  10.1038/s41598-018-20539-2}{Sci. Rep. \textbf{8}, 2973 (2018)}], the section on ``Absence of dc nonreciprocal current in noninteracting systems'' contains an incorrect argument around Eq.~(14). Namely, the nonreciprocal current proportional to $E^2$ may exist in time reversal symmetric noninteracting systems in general. Such nonreciprocal current $\propto E^2$ can be studied  based on the Keldysh Green's function method developed in this paper, which would be an interesting future problem.
}. 

\begin{figure}
\centering{}\includegraphics[width=1\linewidth]{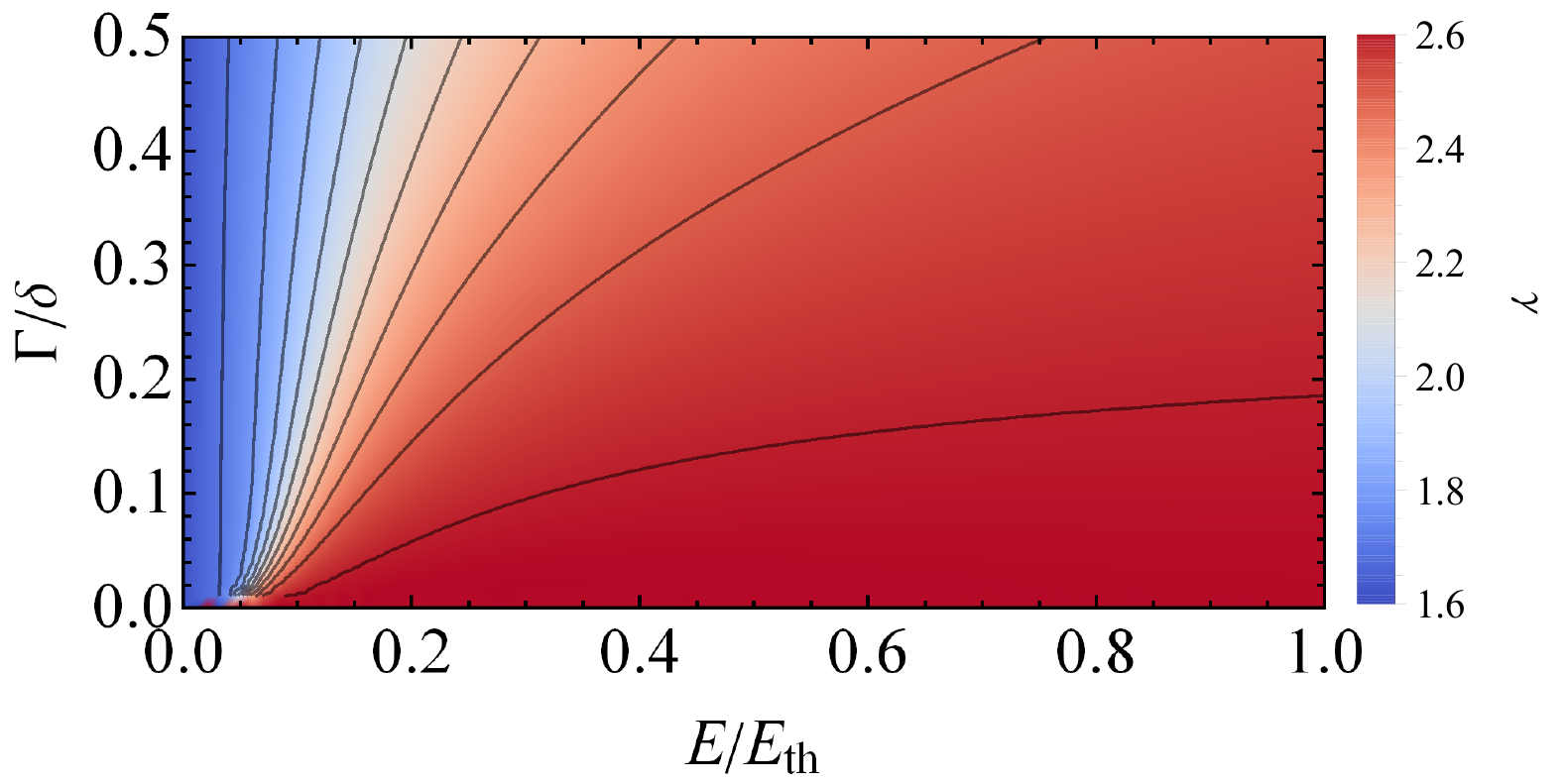}\caption{\label{fig:nrratio}Nonreciprocity ratio $\gamma=J(+E)/J(-E)$ in the steady state for a two band model of noncentrosymmetric insulator with a nonzero shift vector, Eq.~(\ref{eq:ham-rm}). 
Nonreciprocity is enhanced for larger electric field $E$ and is suppressed for stronger dissipation $\Gamma$.}
\end{figure}

To demonstrate the crossover, we introduce a model for a noncentrosymmetric insulator
\begin{equation}
H(k)=\delta \sigma_x + m \sqrt{1+ck^2}\sigma_y+v k\sigma_z,\label{eq:ham-rm}
\end{equation}
where the parameter $m$ controls the strength of inversion breaking [$\sigma_xH(k)\sigma_x\neq H(-k)$] which yields a nonzero shift vector.
Note that this model is time-reversal symmetric, $\sigma_x H^\ast(k)\sigma_x=H(-k)$, 
which prohibits nonreciprocal response that arises from asymmetric band structures such as magnetochiral anisotropy~\cite{RikkenNature}.
We show the nonreciprocity ratio $\gamma$ in Fig.~\ref{fig:nrratio} as a function of the electric field $E$ and the dissipation strength $\Gamma$.
We choose $m=0.5\delta$ and $c=0.5v^2/\delta^2$, which leads to $\gamma_P=2.62$, $\sqrt{\gamma_P}=1.62$.
We note that $\gamma_P$ has no dependence on $E$ and $\Gamma$. 
We can see that the nonreciprocity ratio changes from $\sim\sqrt{\gamma_P}$ to $\sim\gamma_P$ as the field strength is increased, which clearly captures the change of the dominant mechanism for the electric current from
the interband to intraband effect.
Namely, for the weak electric field regime, the interband effect is dominant  since the phase coherence between the two bands is important for the current response with a small number of excited electrons.
For the strong electric field regime, in contrast, the intraband effect becomes dominant which means that there appear many tunnel electrons which carries current according to their group velocity.
In addition, Fig.~\ref{fig:nrratio} shows that strong dissipation $\Gamma$ suppresses nonreciprocity. In particular, we find that nonreciprocity in the crossover regime is quickly suppressed by the dissipation. 

\subsubsection{Nonreciprocal spin transport}
It is interesting to investigate a new type of nonreciprocal transports
that is not characterized by the nonreciprocity of the tunneling probability.
The momentum distribution of the excited electrons due to the tunneling
process is highly asymmetric around the gap minimum (only left or
right is occupied according to the sign of the electric field), which
is a peculiar property absent in metallic systems. 

\begin{figure}
\centering{}\includegraphics[width=\linewidth]{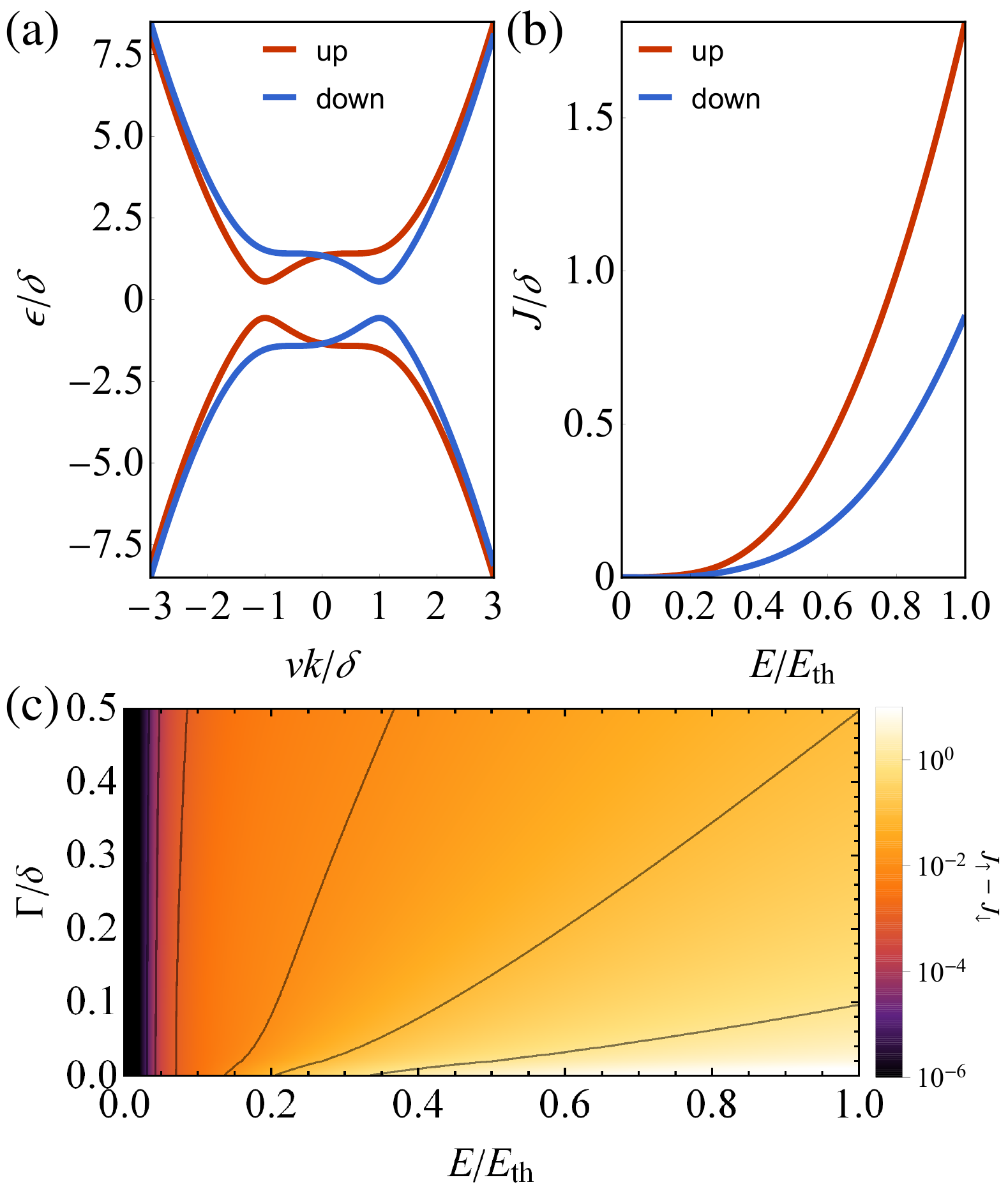}\caption{\label{fig:spin-current}Nonreciprocal spin current in the nonequilibrium steady state of a Rashba-splitted insulator, Eq.~(\ref{eq:spin-accumulation}). $\lambda=0.4\delta,$ $\gamma=1.25v^{2}/\delta$. (a) Energy dispersion. (b) Spin-resolved
current as a function of electric field $E$ for $\Gamma=0.1\delta$. (c) Nonreciprocal spin current $J_\uparrow-J_\downarrow$ against electric field $E$ and dissipation $\Gamma$. $J_\uparrow(-E)=-J_\downarrow(E)$ leads to directionality $J_\uparrow(E)-J_\downarrow(E)=J_\uparrow(-E)-J_\downarrow(-E)$.  }
\end{figure}

We can exploit this feature to obtain a nonreciprocal \textit{spin} transport when
the band dispersion has a skew around the gap minimum. Under the time-reversal
symmetry, however, the gap minimum with an opposite skew exists at
$-k$, so that the asymmetry in the electric current should vanish
if contributions from this pair of gap minima is added up. The nonreciprocal
transport due to this asymmetry may survive when we consider the spin
current. We here consider an insulating model with a Rashba spin-orbit
coupling 
\begin{equation}
H(k)=(vk+\lambda s_{z})\sigma_{x}+(\delta-\gamma k^{2})\sigma_{z},\label{eq:spin-accumulation}
\end{equation}
where $s_{z}$ is the (real) spin of the electron. 
This model is time-reversal symmetric since $\sigma_z s_y H^\ast(k) s_y\sigma_z=H(-k)$, while it lacks the inversion symmetry as $\sigma_z H(k) \sigma_z\neq H(-k)$.
We plot the energy
dispersion of this Hamiltonian in Fig.~\ref{fig:spin-current}(a).
Due to the Rashba spin splitting, time-reversal partner at $-k$ has
the opposite spin polarization. Thus the tunneling current for the
spin up and down differs due to the skewed dispersion, as shown in
Fig.~\ref{fig:spin-current}(b). 
The spin current due to this difference, 
shown in Fig.~\ref{fig:spin-current}(c), does not change when the electric field is inverted, i.e., the spin
current exhibits nonreciprocity. This is a new type of nonreciprocal
transports which is absent in the metallic transport with the shift
of the Fermi surface. 
Note that there are two pairs of saddle points for each spin sector of this model, 
and we have neglected the pair with larger threshold field, for simplicity. We also have neglected a ($E$-dependent) slight deviation of the crossing point $z_1=k-Et_g$ from the gap minimum. 

Recently, spin dependent transport has been found in DNA molecules~\cite{Gohler11}, and spin transport in chiral materials
(chiral-induced spin selectivity (CISS)) is attracting growing interests ~\cite{Matityahu16}. 
In CISS, photoexcited electrons propagate through insulating DNA molecules and show spin accumulation due to spin dependent decay rates. 
Similarly, the above-mentioned spin transport in the tunneling process indicates a spin rectification effect, and can induce spin accumulation in noncentrosymmetric/chiral semiconductors with application of electric fields.
While the present mechanism of spin accumulation applies for tunneling electrons and not for photoexcited electrons in CISS, these two effects could be related with each other in that both induce spin accumulation via electron propagation through an insulator. In particular, the spin current in tunneling problem implies that application of strong dc electric fields to chiral molecules including DNAs can induce spin current generation and spin accumulation.

\subsection{Extension to lattice systems\label{subsec:Extension-to-lattice-system}}

So far, we have considered models in a continuous limit, such as the
Landau-Zener model. Here we briefly introduce an extension of the
formalism to lattice systems with a Brillouin zone. In isolated
lattice systems, the electron passes through the gap minimum periodically,
with the period of the Bloch oscillation $T_{B}=2\pi/|E|a_{0}$ ($a_{0}$
is the lattice constant). Thus the asymptotic form of the tunneling
amplitude $a_{+}(t)$ is modified from Eq.~(\ref{eq:ap1st-ddp})
to 
\begin{equation}
a_{+}(t)\sim\sqrt{P_{0}}\sum_{n=-N}^{\infty}e^{in\int_{0}^{T_{B}}dt(\Delta+ER)}\Theta(t-nT_{B}).\label{eq:ap1st-ddp-lattice}
\end{equation}
Here, $N\rightarrow\infty$ should be taken after the calculation
of Green's functions for the open system, to avoid the divergence
of the sum. 
The $n$ summation appears due to the contribution from the multiple saddle points, which has a phase difference originating from the dynamical phase factor ($W(t+T_B)=W(t)e^{i\int_0^{T_B}dt(\Delta+ER)}$). 

By repeating the derivation in the previous sections with
Eq.~(\ref{eq:ap1st-ddp-lattice}) instead of Eq.~(\ref{eq:ap1st-ddp}), 
one can show that the correction to the nonequilibrium Green's function is modified as
\begin{align}
\left[G_{\text{LZ}}^{<}(t,t)\right]_{\pm\pm} & \rightarrow\dfrac{(1+e^{-2\Gamma T_{B}})\left[G_{\text{LZ}}^{<}(t,t)\right]_{\pm\pm}}{|1-e^{-2\Gamma T_{B}-i\int_{0}^{T_{B}}dt(\Delta+ER)}|^{2}},\\
\left[G_{\text{LZ}}^{<}(t,t)\right]_{+-} & \rightarrow\dfrac{\left[G_{\text{LZ}}^{<}(t,t)\right]_{+-}}{1-e^{-2\Gamma T_{B}-i\int_{0}^{T_{B}}dt(\Delta+ER)}},
\end{align}
for $t\in[0,T_{B})$. The expression for an arbitrary time can be
obtained by employing the periodicity $[G_{\text{LZ}}^{<}(t+T_{B},t+T_{B})]_{\pm\pm}=[G_{\text{LZ}}^{<}(t,t)]_{\pm\pm}$
and $[G_{\text{LZ}}^{<}(t+T_{B},t+T_{B})]_{+-}=[G_{\text{LZ}}^{<}(t,t)]_{+-}e^{i\int_{0}^{T_{B}}dt(\Delta+ER)}$.

\begin{figure}
\centering{}\includegraphics[width=0.85\linewidth]{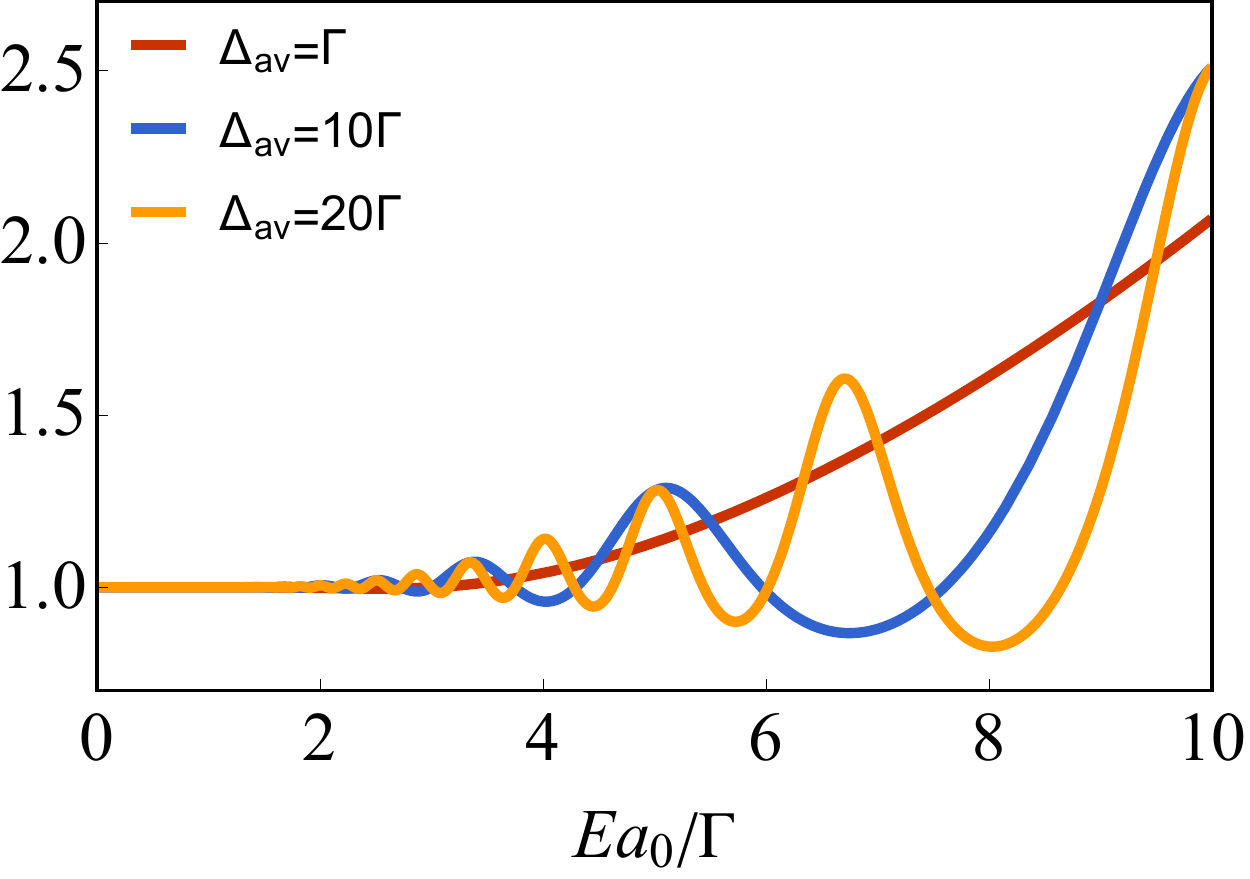}\caption{\label{fig:interference}The interference factor $(1+e^{-2\Gamma T_{B}})/|1-e^{-2\Gamma T_{B}-i\int_{0}^{T_{B}}dt\Delta}|^{2}$
for various values of the momentum average of the energy gap $\Delta_{\text{av}}\coloneqq a_{0}\int_{0}^{2\pi/a_{0}}dk/(2\pi)\times(\varepsilon_{+}-\varepsilon_{-})$,
where $T_{B}=2\pi/|E|a_{0}$. }
\end{figure}

The additional factor characterized by the dynamical phase and $\Gamma T_{B}=2\pi\Gamma/Ea_{0}$
describes the interference between tunneling processes with different
times. The electron excited at $t=nT_{B}$ acquires the dynamical
phase $i\int_{0}^{T_{B}}dt(\Delta+ER)$ relative to the electron excited
at $t=(n+1)T_{B}$. The interference becomes significant when the
electric field is so large that the relaxation time $1/\Gamma$ leading
to the decay of the amplitude is comparable to the period of the tunneling
processes $T_{B}$. We plot the interference factor $(1+e^{-2\Gamma T_{B}})/|1-e^{-2\Gamma T_{B}-i\int_{0}^{T_{B}}dt\Delta}|^{2}$
in Fig.~\ref{fig:interference}. 

\section{Conclusion\label{sec:Conclusion}}

In this paper, we studied the nonequilibrium steady state of the insulating
systems with the nonperturbative correction derived from the quantum
tunneling. We established a new framework for the nonequilibrium Green's function in the tunneling problem, where the Green's function in the
snapshot basis is represented by the solution to the time evolution
of the isolated system that the conventional approaches  are based on.  
We perform an asymptotic evaluation of the nonequilibrium
Green's function in the snapshot basis, which reproduces the result
of the Boltzmann equation with the relaxation-time approximation in
the adiabatic limit. By combining the Lefschetz thimble method, we
also obtain the nonperturbative correction to the nonequilibrium Green's
function, and discuss the electric current in the nonequilibrium steady
state. We also discuss the nonreciprocal transport associated with
the tunneling current, and propose new phenomena, i.e., the crossover
of the nonreciprocity ratio in the nonmagnetic noncentrosymmetric
insulators, and a nonreciprocal spin current derived from the asymmetric
band dispersion in spin-splitted insulators. 

The application
of the present formalism in the strong-field regime turned out to be unexpectedly
successful for the Landau-Zener model.
This should be attributed to
the fact that the asymptotic evaluation of the tunneling probability
coincides with the exact solution. 
Such feature is absent in generic
models (in particular for lattice models with an energy cutoff), and
we have to substantially improve the asymptotic method adopted in the
present study, e.g., by a more sophisticated treatment of the Lefshetz
thimble. Extension of the present formalism to many-body systems~\cite{Oka2003,Sugimoto2008} is
also an important open problem.

\begin{acknowledgements}
We thank Akira Furusaki for fruitful discussions.
This work was supported by JSPS KAKENHI (20K14407) and JST CREST (JPMJCR19T3). TM acknowledges supports from The University of Tokyo Excellent Young Researcher Program, and JST PRESTO (JPMJPR19L9). NN is supported by JST CREST Grant Number JPMJCR1874 and JPMJCR16F1, Japan, and JSPS KAKENHI Grant numbers 18H03676 and 26103006.
\end{acknowledgements}

\appendix

\section{Derivation of the nonequilibrium Green's functions in the snapshot
basis\label{sec:Derivation-of-the}}

Here, we derive the expressions for the nonequilibrium Green's function
in the snapshot basis. Let us begin with the Heisenberg equation of
the annihilation operators,
\begin{align}
i\dot{\hat{c}}_{k\sigma}(t) & =\sum_{\sigma^{\prime}}\langle\sigma|H(k-Et)|\sigma^{\prime}\rangle\hat{c}_{k\sigma^{\prime}}(t)+\sum_{p}V_{p}^{\ast}\hat{b}_{k\sigma p}(t),\\
i\dot{\hat{b}}_{k\sigma p}(t) & =\omega_{p}\hat{b}_{k\sigma p}(t)+V_{p}\hat{c}_{k\sigma}(t).
\end{align}
The latter one can be solved w.r.t. $\hat{b}$ as 
\begin{equation}
\hat{b}_{k\sigma p}(t)=\hat{b}_{k\sigma p}(t_i)e^{-i\omega_{p}(t-t_i)}-iV_{p}\int_{t_i}^{t}dt^{\prime}\hat{c}_{k\sigma}(t^{\prime})e^{-i\omega_{p}(t-t^{\prime})},
\end{equation}
where $t_i=-\infty$ is the initial time where the system is in equilibrium.
By substituting this into the former equation of motion, we obtain
\begin{align}
i\dot{\hat{c}}_{k\sigma}(t) & =\sum_{\sigma^{\prime}}\langle\sigma|H(k-Et)|\sigma^{\prime}\rangle\hat{c}_{k\sigma^{\prime}}(t)+\sum_{p}V_{p}^{\ast}\hat{b}_{k\sigma p}(t_i)e^{-i\omega_{p}(t-t_i)}\nonumber \\
 & -i\sum_{p}|V_{p}|^{2}\int_{t_i}^{t}dt^{\prime}\hat{c}_{k\sigma}(t^{\prime})e^{-i\omega_{p}(t-t^{\prime})}.
\end{align}
The memory effect described by the last term vanishes (i.e., dynamics becomes Markovian) when the fermionic reservoir satisfies the broadband condition Eq.~(\ref{eq:markov}):
The last term is shown to be instantaneous as
\begin{align}
 & \sum_{p}|V_{p}|^{2}\int_{t_i}^{t}dt^{\prime}\hat{c}_{k\sigma}(t^{\prime})e^{-i\omega_{p}(t-t^{\prime})}\nonumber \\
 & =\int d\omega\sum_{p}|V_{p}|^{2}\delta(\omega-\omega_{p})\int_{t_i}^{t}dt^{\prime}\hat{c}_{k\sigma}(t^{\prime})e^{-i\omega(t-t^{\prime})}\\
 & =\int\dfrac{d\omega}{2\pi}2\Gamma\int_{t_i}^{t}dt^{\prime}\hat{c}_{k\sigma}(t^{\prime})e^{-i\omega(t-t^{\prime})}=\Gamma\hat{c}_{k\sigma}(t).\label{eq:markov-trick}
\end{align}
Namely, we obtain 
\begin{align}
i\dot{\hat{c}}_{k\sigma}(t) & =\sum_{\sigma^{\prime}}\langle\sigma|H(k-Et)|\sigma^{\prime}\rangle\hat{c}_{k\sigma^{\prime}}(t)-i\Gamma\hat{c}_{k\sigma}(t)\nonumber \\
 & +\sum_{p}V_{p}^{\ast}\hat{b}_{k\sigma p}(t_i)e^{-i\omega_{p}(t-t_i)}.
\end{align}
Then, by performing the unitary transformation Eq.~(\ref{eq:snapshot-expansion}),
we obtain
\begin{align}
i\dfrac{d}{dt}\begin{pmatrix}\hat{\psi}_{+,k}(t)\\
\hat{\psi}_{-,k}(t)
\end{pmatrix} & =\begin{pmatrix}-i\Gamma & W(t)\\
W^{\ast}(t) & -i\Gamma
\end{pmatrix}\begin{pmatrix}\hat{\psi}_{+,k}(t)\\
\hat{\psi}_{-,k}(t)
\end{pmatrix}\nonumber \\
 & +\sum_{p\sigma}V_{p}^{\ast}\begin{pmatrix}\langle\psi_{+,k}(t)|\sigma\rangle\\
\langle\psi_{-,k}(t)|\sigma\rangle
\end{pmatrix}\hat{b}_{k\sigma p}(t_i)e^{-i\omega_{p}(t-t_i)}.\label{eq:eom-psi}
\end{align}

In order to solve this differential equation, we introduce the unitary
matrix $U(t)$ defined as Eq.~(\ref{eq:unitary}). By replacing the
offdiagonal matrix in the right-hand side as 
\begin{equation}
\begin{pmatrix}0 & W(t)\\
W^{\ast}(t) & 0
\end{pmatrix}=i\dot{U}(t)U^{\dagger}(t),
\end{equation}
we can deform Eq.~(\ref{eq:eom-psi}) into
\begin{multline}
i\dfrac{d}{dt}\left[U^{\dagger}(t)\begin{pmatrix}\hat{\psi}_{+,k}(t)\\
\hat{\psi}_{-,k}(t)
\end{pmatrix}e^{\Gamma t}\right]\\
=\sum_{p\sigma}V_{p}^{\ast}U^{\dagger}(t)\begin{pmatrix}\langle\psi_{+,k}(t)|\sigma\rangle\\
\langle\psi_{-,k}(t)|\sigma\rangle
\end{pmatrix}\hat{b}_{k\sigma p}(t_i)e^{-i\omega_{p}(t-t_i)+\Gamma t},
\end{multline}
which we can solve just by integrating on $[t_i,t]$.
Especially,
when $\Gamma=V_{p}=0$ (i.e., the case of the isolated system), we
obtain 
\begin{equation}
\begin{pmatrix}\hat{\psi}_{+,k}(t)\\
\hat{\psi}_{-,k}(t)
\end{pmatrix}=U(t)U^{\dagger}(t_i)\begin{pmatrix}\hat{\psi}_{+,k}(t_i)\\
\hat{\psi}_{-,k}(t_i)
\end{pmatrix},
\end{equation}
by which the expression for $[G_{0}^{R}(t,t^{\prime})]_{\alpha\beta}=-i\langle\{\hat{\psi}_{\alpha,k}(t),\psi_{\beta,k}^{\dagger}(t^{\prime})\}\rangle_{0}\Theta(t-t^{\prime})$,
Eq.~(\ref{eq:G0R}), immediately follows. When $\Gamma\neq0$,
$V_{p}\neq0$, we arrive at
\begin{multline}
\begin{pmatrix}\hat{\psi}_{+,k}(t)\\
\hat{\psi}_{-,k}(t)
\end{pmatrix}
= iG_0^R(t,t_i)\begin{pmatrix}\hat{\psi}_{+,k}(t_i)\\
\hat{\psi}_{-,k}(t_i)
\end{pmatrix}e^{-\Gamma (t-t_i)}\\
+\int_{-\infty}^{\infty}d\tau \sum_{p\sigma}V_{p}^{\ast}G_0^R(t,\tau)\begin{pmatrix}\langle\psi_{+,k}(\tau)|\sigma\rangle\\
\langle\psi_{-,k}(\tau)|\sigma\rangle
\end{pmatrix}\\\times\hat{b}_{k\sigma p}(t_i)e^{-i\omega_{p}(\tau-t_i)-\Gamma (t-\tau)}.
\end{multline}
where the first term vanishes in $t_i\rightarrow-\infty$.
Now the field operator $\hat{\psi}$ is expressed by the bath operator
$\hat{b}$ at the infinite past. As the bath fermions are in equilibrium
at the infinite past, we can evaluate the Green's functions of $\hat{\psi}$
by using 
\begin{gather}
\{\hat{b}_{k\sigma p}^{\dagger}(t_i),\hat{b}_{k\sigma^{\prime}q}(t_i)\}=\delta_{\sigma\sigma^{\prime}}\delta_{pq},\\
\langle\hat{b}_{k\sigma p}^{\dagger}(t_i)\hat{b}_{k\sigma^{\prime}q}(t_i)\rangle=\delta_{\sigma\sigma^{\prime}}\delta_{pq}f_{D}(\omega_{p}).
\end{gather}
For the retarded Green's function, one can derive Eq.~(\ref{eq:GR}) as
\begin{align}
G^{R}(t,t^{\prime}) & =2\Gamma\int_{-\infty}^{t^{\prime}}d\tau G_{0}^{R}(t,t^{\prime})e^{-\Gamma(t+t^{\prime}-2\tau)}\\
 & =G_{0}^{R}(t,t^{\prime})e^{-\Gamma(t-t^{\prime})},
\end{align}
by using
$\sum_{p}|V_{p}|^{2}e^{-i\omega_{p}(\tau-\tau^{\prime})}=2\Gamma\delta(\tau-\tau^{\prime})$
(See Eq.~(\ref{eq:markov-trick})), $\sum_\sigma\langle\psi_{\alpha,k}(\tau)|\sigma\rangle\langle\sigma|\psi_{\beta,k}(\tau)\rangle=\delta_{\alpha\beta}$ and  $G_{0}^{R}(t,\tau)[G_{0}^{R}(t^{\prime},\tau)]^{\dagger}=iG_{0}^{R}(t,t^{\prime})\Theta(t^{\prime}-\tau)$ for $t>t^{\prime}$. 
The expression for the lesser component, Eqs.~(\ref{eq:Glesser}), (\ref{eq:Sigma-lesser}), can also be derived using  
\begin{multline}
\sum_{p}|V_{p}|^{2}\langle\hat{b}_{k\sigma p}^{\dagger}(t_i)\hat{b}_{k\sigma p}(t_i)\rangle e^{-i\omega_{p}(\tau-\tau^{\prime})}\\=2\Gamma\int\dfrac{d\omega}{2\pi}f_{D}(\omega)e^{-i\omega(\tau-\tau^{\prime})}.
\end{multline}

\section{Nonperturbative contribution to the drift effect\label{sec:Nonperturbative-contribution-to}}

As can be seen in Fig.~\ref{fig:P-adiabatic}~(b),
the drift effect described by the relaxation-time approximation,
Eq.~(\ref{eq:boltzmann}), overestimates the height of the peak. 
This is due to the nonperturbative effect nonnegligible around the band
top.

The failure of the approximation is derived from the order-by-order
evaluation of the gradient expansion, formally expressed by the exponential operator $\exp(-\partial_\Gamma\partial_\tau/2)$ in Eq.~(\ref{eq:g-ad-partial}).
The exact result is recovered by replacement of the expression, 
\begin{equation}
e^{-\partial_{\Gamma}\partial_{\tau}/2}\left.\dfrac{\varepsilon(t-\tau)-\varepsilon(t)}{2\Gamma}\right|_{\tau=0}\rightarrow\int_{0}^{\infty}d\tau e^{-2\Gamma\tau}(\varepsilon(t-\tau)-\varepsilon(t)).
\end{equation}

Let us consider to apply the saddle point (thimble) method to the
$\tau$ integral. The saddle point must satisfy
\begin{equation}
-2\Gamma+\dfrac{\partial_{t}\varepsilon(t-\tau)}{\varepsilon(t)-\varepsilon(t-\tau)}=0.\label{eq:saddle-drift}
\end{equation}
When $\tau$ is so small that we can approximate the denominator by
$\tau\partial_{t}\varepsilon(t-\tau)$, we obtain $\tau=1/2\Gamma$,
which recovers the result of the gradient expansion at the first order. 

On the other hand, when $\varepsilon(t)=\varepsilon(-t)$ holds,
$\tau=2t+1/2\Gamma$ is also an approximate solution:
\begin{equation}
\dfrac{\partial_{t}\varepsilon(t-(2t+1/2\Gamma))}{\varepsilon(t)-\varepsilon(t-(2t+1/2\Gamma))}=\dfrac{-\partial_{t}\varepsilon(t+1/2\Gamma)}{\varepsilon(t)-\varepsilon(t+1/2\Gamma)}\simeq2\Gamma.
\end{equation}
While the contribution from this saddle point is negligible for large
$t$ due to the factor of $e^{-4\Gamma t}$, it can be relevant when
$t$ is small, i.e., when the tunneling process occurs. 

\begin{figure}
\centering{}\includegraphics[width=0.85\linewidth]{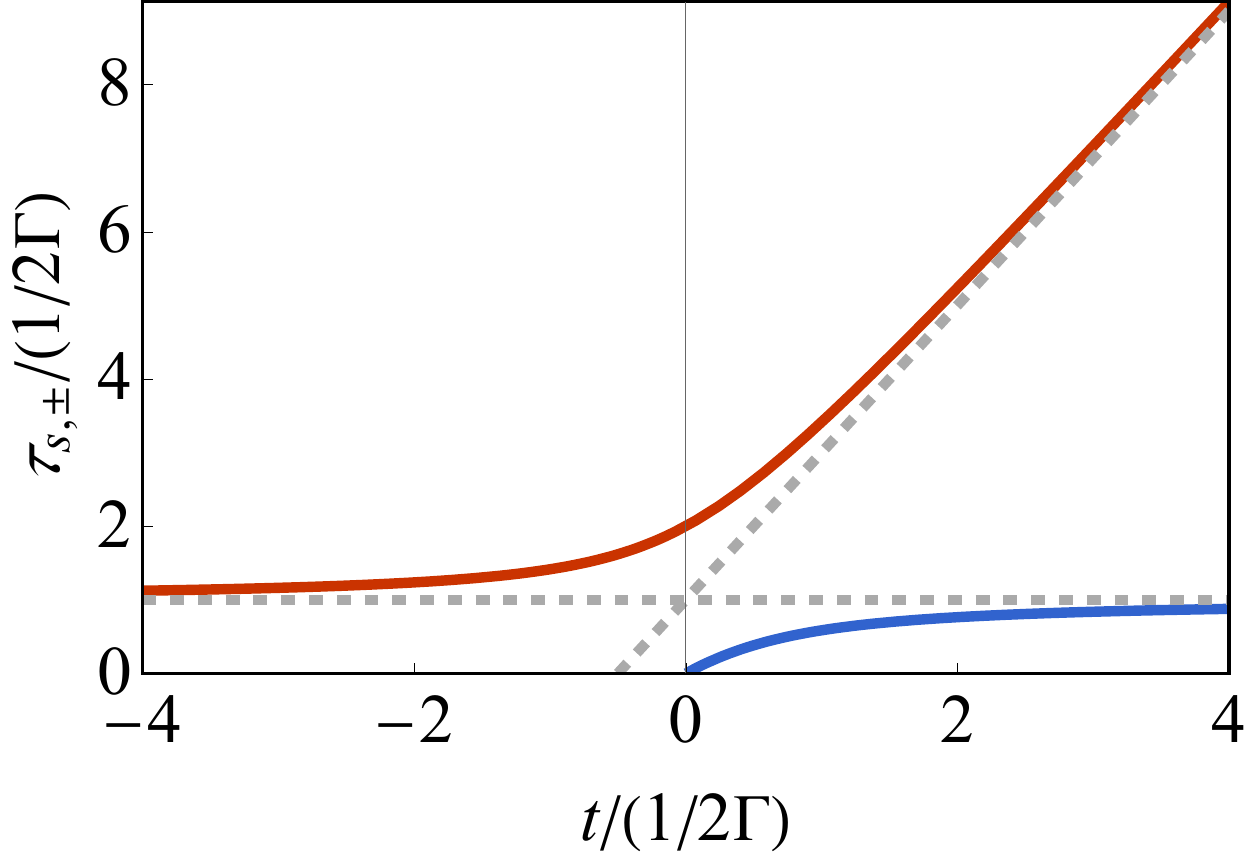}\caption{\label{fig:tau}The position of the saddle point $\tau_{s,\pm}$ (red for $+$ and blue for $-$) as a function of $t$. Dashed lines are $\tau=1/2\Gamma, 2t+1/2\Gamma$.}
\end{figure}

Let us see this contribution from the additional saddle point using
a specific example. For the Landau-Zener model, $\varepsilon(t)=\sqrt{(vEt)^{2}+\delta^{2}}$,
Eq.~(\ref{eq:saddle-drift}) reads
\begin{equation}
(t-\tau)^{2}+\dfrac{t-\tau}{\Gamma}-t^{2}+\dfrac{1}{4\Gamma^{2}}\dfrac{(t-\tau)^{2}}{(t-\tau)^{2}+(\delta/vE)^{2}}=0.
\end{equation}
While this is a quartic equation, the last term can be neglected regardless
of the value of $t-\tau$, if the adiabatic condition $1/\Gamma\ll\delta/vE$
is satisfied. We obtain
\begin{equation}
\tau_{s,\pm}\simeq t+\dfrac{1}{2\Gamma}\pm\sqrt{t^{2}+\dfrac{1}{4\Gamma^{2}}}
\end{equation}
for the approximate position of the saddle point.
As we show in Fig.~\ref{fig:tau}, $\tau_{s,\pm}$ is 
significantly deviated from $\tau=1/2\Gamma,2t+1/2\Gamma$ around the gap minimum $t=0$, 
in addition to the fact that both of the two saddle points are relevant here.
By applying the saddle point method, we obtain
\begin{equation}
2\Gamma\int_{0}^{\infty}d\tau e^{-2\Gamma\tau}(\varepsilon(t-\tau)-\varepsilon(t))\simeq\dfrac{E}{2\Gamma}(v_{+}(t)+v_{-}(t)\Theta(t)),
\end{equation}
where
\begin{align}
v_{\pm}(t) & =\dfrac{\sqrt{2\pi}}{e}\dfrac{\partial\varepsilon}{\partial k}\left(\sqrt{\dfrac{(\sqrt{4\Gamma^{2}t^{2}+1}\pm1)^{3}}{(2\Gamma)^{2}\sqrt{4\Gamma^{2}t^{2}+1}}}\right)e^{-(2\Gamma t\pm\sqrt{4\Gamma^{2}t^{2}+1})}.
\end{align}
We plot this result by a green line in Fig.~\ref{fig:P-adiabatic},
which accurately follows the numerical result.

\section{Tunneling amplitude evaluated by the Lefschetz thimble method\label{sec:Tunneling-amplitude-evaluated}}

In this appendix, we explain how to calculate the asymptotic form
of the tunneling amplitude $a_{+}(t)$, Eq.~(\ref{eq:ap1st}), using the Lefschetz thimble
method~\cite{Fukushima2020}.

\subsection{Analytic continuation}

First, we perform the analytic continuation of the integrand
to rewrite Eq.~(\ref{eq:ap1st}) as a contour
integral in the complex plane. We here introduce a complexified momentum
$k-Et_1\rightarrow z_1\in\mathbb{C}$ (and $k-Et^{\prime}\rightarrow z^{\prime}\in\mathbb{C}$ in the phase factor)
as the variable of integration.

We note that, in analytic continuation, we have to be careful on the
treatment of the Berry connection difference $A_{++}-A_{--}$ in the
phase factor of Eq.~(\ref{eq:Wt}), which is not gauge-invariant and not necessarily
analytic. It is convenient to employ the alternative expression, Eq.~(\ref{eq:Wt-R}), 
for the integrand $W(t)$ with the shift vector $R=A_{++}-A_{--}-\partial_{k}\arg A_{+-}$ 
to circumvent this problem. 
This expression is analytic w.r.t. $k-Et_1$ in generic cases.

To avoid confusion, let us introduce $\tilde{A}_{+-}(z_1)$ and $\tilde{R}(z_1)$
as an analytic continuation of $|A_{+-}(k-Et_1)|$ and $R(k-Et_1)$, respectively.
Then Eq.~(\ref{eq:ap1st}) reads
\begin{equation}
a_{+}(t)=ie^{i\arg A_{+-}(0)}\int_{C_0}dz_1\tilde{A}_{+-}(z_1)e^{-i\int_{0}^{z_1}dz^{\prime}(\Delta/E+\tilde{R})},\label{eq:ap1st-cntr}
\end{equation}
where $\Delta\coloneqq\varepsilon_{+}-\varepsilon_{-}$. $C_0$ denotes the half line on the real axis, $z_1=x\in\mathbb{R},\,x:\text{sgn}(E)\times \infty\rightarrow k-Et$.

There are exceptional cases where $|A_{+-}(k-Et_1)|$ and $R(k-Et_1)$ cannot be analytically continued.
Such a situation happens when there exists a gauge choice such that $A_{++}(k)=A_{--}(k)$ and $A_{+-}(k)\in\mathbb{R}$ with $A_{+-}({}^\exists k_{a})=0$ hold,
because the shift vector becomes $R(k)=\pi\sum_{k_{a}}\delta(k-k_{a})$
(mod $2\pi$). 
Still, in such cases, the combined quantity $|A_{+-}(k-Et_1)|e^{-i\int_{0}^{k-Et_1}dk^{\prime}R}=\pm A_{+-}(k-Et_1)$
is analytic and gauge-invariant (up to the phase factor $e^{i\arg A_{+-}(0)}$).
Thus, as an exceptional treatment, we introduce $\tilde{A}_{+-}(z_1)$ as an analytic continuation of $A_{+-}(k-Et_1)$ in the above-mentioned gauge instead,
and set $\tilde{R}(z_1)=0$.

\subsection{Analytic property of $2\times2$ Hamiltonian}

When the system is described by a $2\times2$ Hamiltonian, one can
express the Hamiltonian using a pseudospin $\bm{\sigma}$ as 
\begin{equation}
H(k)=d_{0}(k)I_{2\times2}+\bm{d}(k)\cdot\bm{\sigma},
\end{equation}
with $\bm{\sigma}$ being the Pauli matrices. We assume that $\bm{d}$
is an analytic function of $k$. Then the analytically-continued variables
are expressed as~\cite{Kitamura2020}
\begin{gather}
\Delta(z_1)=2\sqrt{\bm{d}^{2}},\\
\tilde{A}_{+-}(z_1)=\dfrac{\sqrt{(\bm{d}\times\partial_{k}\bm{d})^{2}}}{2\bm{d}^{2}},\\
\tilde{R}(z_1)=\dfrac{(\bm{d}\times\partial_{k}\bm{d})\cdot\partial_{k}^{2}\bm{d}}{(\bm{d}\times\partial_{k}\bm{d})^{2}}\sqrt{\bm{d}^{2}}.
\end{gather}
Note that this expression includes the exceptional cases mentioned in the previous subsection, which correspond to the situation where $(\bm{d}\times\partial_{k}\bm{d})\cdot\partial_{k}^{2}\bm{d}\equiv0$.
Because $\tilde{R}$ is indeterminate at $k=k_a$ with $(\bm{d}\times\partial_{k}\bm{d})^{2}|_{k=k_a}=0$, $\tilde{R}$ can be a singular function when the branch of $\sqrt{(\bm{d}\times\partial_{k}\bm{d})^{2}}$ for $\tilde{A}_{+-}$ is not appropriately chosen.

As the gap closing point $z_1=k_{c}$ with $\Delta(k_{c})=0$ plays
a key role below, let us see properties of the above variables in the
vicinity of $z_1=k_{c}$. 
The gap closing points appear in a pairwise manner (i.e., $\Delta(k_c)=\Delta(k_c^\ast)=0$), because $\bm{d}(z_1^{\ast})=[\bm{d}(z_1)]^{\ast}$
holds for Hermitian Hamiltonian $\bm{d}(k\in\mathbb{R})\in\mathbb{R}^{3}$.
For future convenience, we label the gap closing points as $k_c^{(\pm1)}, k_c^{(\pm2)}, \dots$ with  $k_c^{(-n)}\coloneqq(k_c^{(n)})^\ast$.

Since $\bm{d}^{2}$ is analytic, $\bm{d}^{2}$
should be expanded as $\bm{d}^{2}=\alpha_{1}^{(n)}(z_1-k_{c}^{(n)})+\alpha_{2}^{(n)}(z_1-k_{c}^{(n)})^{2}+\dots$,
with $\alpha_{1}^{(n)}\neq0$ for generic cases. Namely, the gap closing
point behaves as a square-root branch point 
\begin{equation}
\Delta(z_1)\sim2\sqrt{\alpha_{1}^{(n)}(z_1-k_{c}^{(n)})}\label{eq:gap-kc}.
\end{equation}

In a similar way, we assume that $(\partial_{k}\bm{d})^{2}=\beta_{0}^{(n)}+\beta_{1}^{(n)}(z_1-k_{c}^{(n)})+\dots$
and $(\bm{d}\times\partial_{k}\bm{d})\cdot\partial_{k}^{2}\bm{d}=\eta_{0}^{(n)}+\eta_{1}^{(n)}(z_1-k_{c}^{(n)})+\dots$
with $\eta_{0}^{(n)}\neq0$. Then, we obtain 
\begin{align}
(\bm{d}\times\partial_{k}\bm{d})^{2} & =\bm{d}^{2}(\partial_{k}\bm{d})^{2}-\dfrac{1}{4}(\partial_{k}\bm{d}^{2})^{2}\\
 & =-\dfrac{1}{4}\alpha_{1}^{(n)2}+\alpha_{1}^{(n)}(\beta_{0}^{(n)}-\alpha_{2}^{(n)})(z_1-k_{c}^{(n)})+\dots,
\end{align}
which leads to
\begin{gather}
\tilde{A}_{+-}(z_1)\sim\dfrac{\zeta_n\text{sgn}(\text{Im}k_c^{(n)})}{4i(z_1-k_{c}^{(n)})},\label{eq:apm-kc}\\
\tilde{R}(z_1)\sim-\dfrac{4\eta_{0}^{(n)}}{\alpha_{1}^{(n)3/2}}\sqrt{z_1-k_{c}^{(n)}},
\end{gather}
as leading-order expressions. $\zeta_n=\zeta_{-n}=\pm1$ arises from the multivaluedness of  $\sqrt{(\bm{d}\times\partial_{k}\bm{d})^{2}}$.

\subsection{Saddle points}

In order to apply the Lefschetz thimble method to the evaluation of Eq.~(\ref{eq:ap1st-cntr}), we need to identify the position
of the saddle point of $f(z_1)$ with $a_{+}(t)=ie^{i\arg A_{+-}(0)}\int_{C_0}dz_1e^{f(z_1)}.$
The saddle point is given as the solution of $\partial_{z_1}f(z_1)=0$,
i.e., it satisfies [See Eq.~(\ref{eq:ap1st-cntr})] 
\begin{equation}
\frac{\partial}{\partial z_1}\ln\tilde{A}_{+-}(z_1)-i\dfrac{\Delta(z_1)}{E}-i\tilde{R}(z_1)=0.\label{eq:saddle}
\end{equation}

For simplicity, we focus on $\tilde{R}=0$ cases here. Results for $\tilde{R}\neq0$  can be recovered by replacing $\Delta$ by $\Delta+E\tilde{R}$ in the final expression (See Ref.~\cite{Kitamura2020} for details). 

For now, we consider $E>0$. Since the second
term diverges in $E\rightarrow0$, the saddle points approach the gap closing point $k_{c}$. However, the first term also diverges in this limit, since 
\begin{equation}
\frac{\partial}{\partial z_1}\ln\tilde{A}_{+-}(z_1)\sim-\frac{\partial}{\partial z_1}\ln(z_1-k_{c})=-\dfrac{1}{z_1-k_{c}}
\end{equation}
follows from Eq.~(\ref{eq:apm-kc}).
Combined with Eq.~(\ref{eq:gap-kc}), the solutions of Eq.~(\ref{eq:saddle}), $z_1=k_s$, at the leading order of $E$ are given as
\begin{equation}
k_{s}^{(n,m)}-k_{c}^{(n)}\sim\left(\dfrac{E^2}{4\alpha_{1}^{(n)}}\right)^{1/3}e^{-\pi i+4\pi im/3}
\end{equation}
with $m=0,1,2$. Due to the branch point, $\arg(k_{s}^{(n,m)}-k_{c}^{(n)})$ is mod $4\pi$ here. 
We note that, in contrast to the gap closing point, $z_1=(k_{s}^{(n,m)})^{\ast}$
is not the saddle point.

Let us evaluate the integral along the thimble (steepest descent) $\mathcal{J}_{n,m}$ associated with the saddle point $z_1=k_s^{(n,m)}$.
Since $f(z_1)-f(k_s^{(n,m)})\in\mathbb{R}$ ($z_1\in \mathcal{J}_{n,m}$) takes the maximal value at $z_1=k_s^{(n,m)}$, the integral can be approximated as
\begin{align}
\int_{\mathcal{J}_{n,m}}dz_1e^{f(z_1)} & \sim\int_{\mathcal{J}_{n,m}}dz_1e^{f(k_{s}^{(n,m)})+f^{\prime\prime}(k_{s}^{(n,m)})(z_1-k_{s}^{(n,m)})^{2}/2}
\end{align}
as known as Laplace's method.
Using
\begin{align}
f^{\prime\prime}(k_{s}^{(n,m)})\sim3\left(\frac{2\alpha_{1}^{(n)2}}{E^{4}}\right)^{1/3}e^{-2\pi im/3},
\end{align}
we can parametrize the steepest descent around $z_1=k_s^{(n,m)}$ as $z_1-k_s^{(n,m)}=(\alpha_{1}^{(n)})^{-1/3}xe^{-i\pi/2+4\pi im/3}$ with $x\in\mathbb{R}$.
Here, the direction of the contour around the saddle point $k_s^{(n,m)}$ is counterclockwise seen from the gap closing point $k_c^{(n)}$.
Combined with
\begin{align}
e^{f(k_{s}^{(n,m)})}\sim i\zeta_n\text{sgn}(\text{Im}k_c^{(n)})\left(\dfrac{e^2\alpha_{1}^{(n)}}{16E^2}\right)^{1/3}e^{-4\pi im/3}e^{-i\int_{0}^{k_{c}^{(n)}}dz^{\prime}\Delta/E},
\end{align}
we obtain the asymptotic form of the integral as
\begin{align}
\int_{\mathcal{J}_{n,m}}dz_1e^{f(z_1)}\sim \zeta_n\text{sgn}(\text{Im}k_c^{(n)})\sqrt{\dfrac{\pi}{3}}\dfrac{e^{2/3}}{2}e^{-i\int_{0}^{k_{c}^{(n)}}dz^{\prime}\Delta/E}.\label{eq:thimble}
\end{align}
According to the exact result obtained by the DDP method~\cite{Davis1976},
the prefactor $\sqrt{\pi/3}e^{2/3}/2=0.9965\dots$ should be replaced by unity when the higher-order terms of the adiabatic perturbation theory are taken into account.
Hereafter we drop this prefactor.

When $E<0$, the position of the saddle point around $k_c^{(n)}$ reads
\begin{equation}
k_{s}^{(n,m)}-k_{c}^{(n)}\sim\left(\dfrac{|E|^{2}}{4\alpha_{1}^{(n)}}\right)^{1/3}e^{\pi i-4\pi im/3},
\end{equation}
which corresponds to $(k_s^{(-n,m)})^\ast$ in the $E>0$ case. 
The expression for the integral coincides with Eq.~(\ref{eq:thimble}) (Note that $E$ in the exponent becomes negative). 

\subsection{Tunneling amplitude}

Let us apply the Lefschetz thimble method. 
Using Cauchy's integral theorem, we can deform the contour of the integral $C_0$ to a set of steepest descents~\cite{Witten2010,Fukushima2020}
\begin{gather}
C=\sum_{n,m}N_{n,m}\mathcal{J}_{n,m}-\Gamma(t),
\end{gather}
where the sum of the contour is defined as $\int_{\Gamma_1\pm\Gamma_2}\coloneqq\int_{\Gamma_1}\pm\int_{\Gamma_2}$.
Here, $\Gamma(t)$ represents the steepest descent extending from the end point of the original contour $C_0$, i.e., $z_1=k-Et$. 
The Morse index $N_{n,m}=\langle C_0,\mathcal{K}_{n,m}\rangle\in\{-1,0,1\}$ 
counts the (oriented) number of intersection between the original contour $C_0$ and the steepest ascent $\mathcal{K}_{n,m}$ associated with $k_s^{(n,m)}$. The orientation is defined as $\langle \mathcal{J}_{n,m},\mathcal{K}_{n^\prime,m^\prime}\rangle=\delta_{n,n^\prime}\delta_{m,m^\prime}$.
Namely, if we neglect the contribution from $\Gamma(t)$, We can rewrite Eq.~(\ref{eq:ap1st-cntr}) as a sum of Eq.~(\ref{eq:thimble}),
\begin{align}
a_{+}(t) & \sim i\sum_{n,m}N_{n,m} \zeta_n\text{sgn}(\text{Im}k_c^{(n)})e^{-i\int_{0}^{k_{c}^{(n)}}dz^{\prime}\Delta/E+i\arg A_{+-}(k)}.
\end{align}

The remaining task is to identify the Morse index $N_{n,m}$. 
As the extension to the case of the multiple pairs of gap closing points is straightforward,
here let us assume that $N_{n,m}=\delta_{n,1}\delta_{m,0}N_{1,0}$ holds for $E>0$,
and the thimble $\mathcal{J}_{1,0}$ passes through $z_1=k-Et_g^{(1)}$, i.e., the momentum at $t_1=t_g^{(1)}$. 
Without calculating the steepest descent directly, whether the latter assumption is consistent can be verified by $\text{Re}f(k_s^{(1,0)})<\text{Re}f(k-Et_g^{(1)})$, which must hold since they are on the same steepest ascent $\mathcal{K}_{1,0}$. The position of $z_1=k-Et_g^{(1)}$ can also be identified by comparing $\text{Im}f$. Note that, while $z_1=k-Et_g^{(1)}$ coincides with the gap minimum for the Landau-Zener model, it is not necessarily the case for generic models (e.g., Eq.~(\ref{eq:spin-accumulation})). In particular, $t_g$ can be a function of $E$.

As the steepest ascent $\mathcal{K}_{1,0}$ has an intersection with $C_0$ when $t>t_g^{(1)}$ (as $z_1=x\in[k-Et,+\infty)$ for $z_1\in C_0$), the Morse index is given as 
\begin{equation}
N_{1,0}=-\text{sgn}(\text{Im}k_c^{(1)})\Theta(t-t_g^{(1)}).
\end{equation}
Here, the sign factor arises because $C_0$ is clockwise (counterclockwise) seen from the gap closing point $k_c^{(1)}$ in the upper (lower) half plane.

When $E<0$, $N_{n,m}=\delta_{n,-1}\delta_{m,0}N_{-1,0}$ should hold, since $-i\int_{0}^{k_{c}^{(-n)}}dz^{\prime}\Delta/E=[-i\int_{0}^{k_{c}^{(n)}}dz^{\prime}\Delta/|E|]^\ast$. 
Now the original contour is $C_0=(-\infty,k-Et]$, and is counterclockwise (clockwise) seen from $k_c^{(-1)}$ on the upper (lower) half plane.
Namely, \begin{equation}
N_{-1,0}=\text{sgn}(\text{Im}k_c^{(-1)})\Theta(t-t_g^{(1)}).
\end{equation}

We can summarize the above results as
\begin{align}
a_{+}(t) & \sim -i\zeta_1\text{sgn}(E)\sqrt{P_0}\Theta(t-t_g^{(1)}) e^{-i\text{Re}\int_{0}^{k_{c}^{(1)}}dz^{\prime}(\Delta/E+\tilde{R})+i\arg A_{+-}(0)}\label{eq:ap1st-ddp-detail}
\end{align}
where
\begin{align}
P_0&= e^{2\text{Im}\int_{0}^{k_{c}^{(1)}}dz^{\prime}(\Delta/|E|+\text{sgn}(E)\tilde{R})}
\end{align}
is the tunneling probability.

When there is only one pair of the gap closing points ($z=k_c^{(\pm1)}$), we can set $t_g^{(1)}=0$ by choosing $k$ and $A_{+-}(0)$ such that the asymptotic form of the tunneling amplitude is real: $a_{+}(t) \sim \sqrt{P_0}\Theta(t)$. This expression is used in the main text for simplicity. 
We note that in such a case the interband matrix element $W(t)$ reads
\begin{align}
W(t)=i\zeta_1|E||A_{+-}(k-Et)|e^{-i\text{Re}\int_{k_{c}^{(1)}}^{k-Et}dz^{\prime}(\Delta/E+\tilde{R})},\label{eq:Wt-mod}
\end{align}
which is used for the evaluation of the electric current in Sec.~\ref{subsec:nonperturbative-electric} ($\zeta_1=1$ is assumed in the main text).

\section{Evaluation of the time-difference factor in the gradient expansion\label{sec:time-difference}}
In the evaluation of $e^{-\partial_{s}\partial_{\tau}-\partial_{s^{\prime}}\partial_{\tau^{\prime}}}I(s,s^{\prime})$ with $t>t^\prime$, we have to deal with $\langle T(t,t^\prime)|\coloneqq e^{-\partial_{s}\partial_{\tau}}e^{-s(t-t^{\prime})}\langle\overline{\psi}_{\alpha,k}(t,\tau)|F(s)$ with $s=\Gamma+i\varepsilon_{\alpha}(t),\tau=0$, and an arbitrary function $F(s)$.
As we have mentioned in the main text, $e^{-s(t-t^{\prime})}$ acts as a time-translation operator as
\begin{align}
\langle T(t,t^\prime)| & =e^{-s(t-t^{\prime})}e^{-\partial_{s}\partial_{\tau}}e^{(t-t^{\prime})\partial_{\tau}}\langle\overline{\psi}_{\alpha,k}(t,\tau)|F(s)\\
 & =e^{-s(t-t^{\prime})}e^{-\partial_{s}\partial_{\tau}}\langle\overline{\psi}_{\alpha,k}(t,\tau+t-t^{\prime})|F(s).
\end{align}
As $\tau+t-t^{\prime}$ is no longer small, we need to shift the origin time of the slow component. 
Using the definition of the slow component, Eq.~(\ref{eq:slow-component}), we obtain
\begin{align}
\langle T(t,t^\prime)|  & =e^{-s(t-t^{\prime})}e^{-\partial_{s}\partial_{\tau}}e^{i\varepsilon_{\alpha}(t)(\tau+t-t^{\prime})-i\varepsilon_{\alpha}(t^{\prime})\tau}\langle\overline{\psi}_{\alpha,k}(t^{\prime},\tau)|F(s)\\
 & =e^{-\Gamma(t-t^{\prime})}e^{-\partial_{s}\partial_{\tau}}e^{-i(\varepsilon_{\alpha}(t)-\varepsilon_{\alpha}(t^{\prime}))\partial_{s}}\langle\overline{\psi}_{\alpha,k}(t^{\prime},\tau)|F(s),
\end{align}
which can be rewritten as 
$\langle T(t,t^\prime)| =
e^{-\partial_{s}\partial_{\tau}}e^{-\Gamma(t-t^{\prime})}\langle\overline{\psi}_{\alpha,k}(t^{\prime},\tau)|F(s)$
with $s=\Gamma+i\varepsilon_{\alpha}(t^{\prime})$, $\tau=0$.
Using this expression, we obtain $[G_{\text{ad}}^{<}(t,t^{\prime})]_{\alpha\beta}$ as
\begin{multline}
\left[G_{\text{ad}}^{<}(t,t^{\prime})\right]_{\alpha\beta}=i2\Gamma e^{-\Gamma(t-t^{\prime})}
e^{-\partial_{s}\partial_{\tau}-\partial_{s^{\prime}}\partial_{\tau^{\prime}}}\\
\times\dfrac{f_{D}(-is)}{s+s^{\prime}}\langle\overline{\psi}_{\alpha,k}(t^\prime,\tau)|\overline{\psi}_{\beta,k}(t^{\prime},\tau^{\prime})\rangle ,
\end{multline}
evaluated at $s=\Gamma+i\varepsilon_{\alpha}(t^\prime),s^{\prime}=\Gamma-i\varepsilon_{\beta}(t^{\prime}),\tau=\tau^{\prime}=0$.

\bibliographystyle{apsrev4-1}
\bibliography{references}

\end{document}